# On Shock Waves and the Role of Hyperthermal Chemistry in the Early Diffusion of Overdense Meteor Trains

Elizabeth A. Silber[*,1], Wayne K. Hocking[2], Mihai L. Niculescu[3], Maria Gritsevich[4,5,6], Reynold E. Silber[7]

[1]Department of Earth, Environmental and Planetary Science, Brown University, Providence, RI, 02912, USA

[2]Department of Physics and Astronomy, University of Western Ontario, London, Ontario, N6A 3K7 Canada

[3]INCAS - National Institute for Aerospace Research "Elie Carafoli", Flow Physics Department, Numerical Simulation Unit, Bucharest 061126, Romania

[4]Department of Physics, University of Helsinki, Gustaf Hällströmin katu 2a, P.O. Box 64, FI-00014 Helsinki, Finland

[5]Department of Computational Physics, Dorodnicyn Computing Centre, Federal Research Center "Computer Science and Control" of the Russian Academy of Sciences, Vavilova St. 40, 119333 Moscow, Russia

[6]Institute of Physics and Technology, Ural Federal University, 620002 Ekaterinburg, Russia

[7]Department of Earth Sciences, University of Western Ontario, London, Ontario, N6A 3B7, Canada

*Corresponding Author

Elizabeth A. Silber
Department of Earth, Environmental and Planetary Science
Brown University
324 Brook St., Box 1946
Providence, RI, 02912
USA
E-mail: elizabeth_silber [at] brown.edu

**Abstract**

Studies of meteor trails have until now been limited to relatively simple models, with the trail often being treated as a conducting cylinder, and the head (if considered at all) treated as a ball of ionized gas. In this article, we bring the experience gleaned in other fields to the domain of meteor studies, and adapt this prior knowledge to give a much clearer view of the microscale physics and chemistry involved in meteor-trail formation, with particular emphasis on the first 100 or so milliseconds of the trail formation. We discuss and examine the combined physico-chemical effects of meteor-generated and ablationally amplified cylindrical shock waves which appear in the ambient atmosphere immediately surrounding the meteor train, as well as the associated hyperthermal chemistry on the boundaries of the high temperature postadiabatically expanding meteor train. We demonstrate that the cylindrical shock waves produced by overdense meteors are sufficiently strong to dissociate molecules in the ambient atmosphere when it is heated to temperatures in the vicinity of 6,000 K, which substantially alters the considerations of the chemical processes in and around the meteor train. We demonstrate that some ambient $O_2$, along with $O_2$ that comes from the shock dissociation of $O_3$, survives the passage of the cylindrical shock wave, and these constituents react thermally with meteor metal ions, thereby subsequently removing electrons from the overdense meteor train boundary through fast, temperature independent, dissociative recombination governed by the second Damköhler number. Possible implications for trail diffusion and lifetimes are discussed.

**Keywords:** meteorites, meteors, meteoroids, shock waves, Earth

**1. Introduction**

The physics of meteoric phenomena can be divided into three basic components (Dressler 2001). The first two, the dynamics of the meteoroid motion in the atmosphere (e.g. Boyd 2000; Gritsevich 2009), and aspects of the chemical and plasma kinetics of thermalized atoms and molecules deposited in the ambient atmosphere by meteor ablation (e.g. Plane 2012; Plane, Feng & Dawkins 2015) have been subject of numerous studies. The third component, which has not received sufficient attention to date, is concerned with cause and effects of meteor generated shock waves and the closely related small scale physical and chemical processes occurring in, and on the boundary of, the extreme environment of the high temperature adiabatically formed meteor trail in the initial stages of the expansion. This aspect of the physico-chemical evolution of overdense meteor trails (defined shortly) is the focus of this study.

Consequently, the broad aim of this work is to present an overview and examine the role of frequently neglected meteor cylindrical shock waves and the associated hyperthermal chemistry. We discuss the nature of physico-chemical and associated processes behind the potentially rapid and short lasting electron removal from post-adiabatically expanding high temperature overdense meteor train boundaries. As the role of meteor-produced shock waves and hyperthermal chemistry phenomena associated with larger meteors have not been covered to a significant degree in the literature, except a very few selected works (e.g. Menees & Park 1976; Park & Menees 1978; Berezhnoy & Borovička 2010), we readdress that issue here. Hence, in this paper, we also present an extended discussion of the relevant aspects of shock waves, hydrodynamic phenomena and hyperthermal chemistry as they may pertain to the topic of early diffusion of overdense meteors.

This paper is organized as follows: in Section 2, we provide the fundamentals and background pertaining to the evolution and behaviour of overdense meteor trains; in Section 3, we discuss the hyperthermal chemistry, while in Section 4 we consider the dynamic and physico-chemical effects of overdense meteor cylindrical shock waves, including a computational model; and finally, our conclusions are presented in Section 5. Our computational model of meteor atmospheric entry at 80 km is discussed for two different meteoroid sizes, and while being modest in scope, nevertheless provides a detailed overview of the main aspects of the flow regimes, with more detailed discussions presented in the Supplementary Material.

**2. Fundamentals and Background**

**2.1 Physical Processes – Formation of the Hydrodynamic Shielding, Initial Radius and Shocks**

Following the initial sputtering regime (Rogers, Hill & Hawkes 2005), it is possible to recognize three distinct stages of the early evolution of the sporadic overdense meteor train at lower altitudes (below ~100 km), beginning with the initial ablation and shock wave formation, and ending with the ambipolar diffusion and chemical removal of electrons from a thermalizing trail. Figure 1 covers some of the features of the items under discussion, and we will refer to it repeatedly during our discussions.

Meteor trails are classified as underdense, transitionally dense or overdense, depending on their so-called "line density" ($q$), or number of electrons per unit length of the trail. Electron densities in the plane perpendicular to the trail are integrated into the line-density calculation. By standard definition, transitional meteors have line densities in the range $2.4 \cdot 10^{14}$ to $10^{16}$ electrons m$^{-1}$, while underdense and overdense meteors are those on the lower and upper ends, respectively, of the transitionally dense trail densities (McKinley 1961; Poulter & Baggaley 1977, 1978). Here we further describe the overdense meteors as particles with diameter between approximately $4 \cdot 10^{-3}$ m and up to small sized fireballs (the latter size corresponding to or exceeding electron line density of $q \sim 10^{19}$ electrons m$^{-1}$) (e.g. Sugar, 1964).

We will now sequentially discuss the various stages of initial trail formation. In the first stage, meteoroids ablate due to high energy hypervelocity collisions with surrounding atmospheric molecules (Öpik 1958; McKinley 1961; McNeil, Lai & Murad. 1998; William & Murad 2002; Vondrak et al. 2008). The high temperature ablated and ionized meteor atoms and electrons, together with ionized and dissociated atmospheric atoms, explosively form a dynamically stable meteor trail cylindrical volume with an initial radius $r_0$, which is approximated as a quasineutral plasma that can subsequently be observed by meteor radars (McKinley 1961; Baggaley & Fisher 1980; Jones 1995; Räbinä et al. 2016). Here, the term initial radius refers to the half-width of the initial (assumed) Gaussian distribution of the ions (or in the case of radio studies, electrons) that has "instantaneously" and adiabatically formed immediately after the passage of the meteoroid, where the volume density of the free electrons is a function of meteoroid mass, size and ionization coefficient (e.g. Jones 1997; Jones & Halliday 2001; Weryk & Brown 2013). The adiabatic formation of the initial trail with radius $r_0$ is accompanied by turbulence generated in the meteor wake ($T \leq 10{,}000$ K), driven by the local flow field velocity, temperature and density gradients (Lees & Hromas 1962). This process is completed within less than the first millisecond and it takes place after the formation and radial expansion of the cylindrical shock wave that will be discussed shortly. However, the density of ionized atoms and electrons in the meteor trail depends on the ionization coefficient (Kaiser 1953; Weryk & Brown 2012, 2013). The majority of meteor radars, especially the lower power ones, detect electrons from the specularly reflecting meteor train (McKinley 1961; Hocking et al. 2016), while a smaller number of higher power radars can obtain reflections directly from the head and in non-specular mode. For the specularly reflecting scenario, overdense meteor trails in particular can be generally treated as metallic cylinders (Poulter & Baggaley 1977; 1978), due to the high electron density and negative dielectric constant associated with plasma.

Ablated meteoric atoms have velocity dependent kinetic energies that may reach several hundred electron volts (eV) (Baggaley 1980). The energy of collisionally released free electrons approaches several eV (e.g. see Baggaley 1980; Hocking et al. 2016 for discussion). The ion energy is converted to intensive heating of the flow field around the meteoroid and also of the ambient atmosphere. Note that our terminology with respect to the meteor generated shocks attempts to reconcile hypersonic and the early meteor shock wave nomenclature (e.g. Bronshten 1965) in light of differences between meteors and a much slower hypersonic flow associated with typical re-entry vehicles.

It is important to emphasize that prior to the first stage of the meteor train evolution (Figure 1), the onset of hydrodynamic shielding (Popova et al. 2001) at higher altitudes (as a precursor to the appearance of the ablation amplified meteor shock front), greatly affects the consideration of the hypervelocity flow in the front of and around the meteoroid (e.g. see Jenniskens et al. 2000; Gritsevich 2008). Moreover, the formation of the hydrodynamic shielding (sometimes referred to as the vapour cap), whose pressure and density are proportional to the cube of the meteoroid velocity (e.g. Öpik 1958; Bronshten 1983; Jenniskens et al. 2000; Popova et al. 2000; Boyd 2000; Campbell-Brown & Koschny 2004), will alter the flow regime considerations (Boyd 2000; Popova et al. 2001) and Knudsen number (Josyula & Burt 2011), shifting the free-molecular flow to higher altitudes, subsequently resulting in the formation of a meteor shock wave front ($T \gg 10\,000$ K) and a related cylindrical shock wave (Figure 1) at higher altitudes (Jenniskens et al. 2000). This region consists of both reflected atmospheric constituents and collisionally ejected meteor atoms and ions. It also exhibits strong velocity dependent density gradients near the meteoroid (Popova et al. 2000; 2001) and may be more than two orders of magnitude larger than the characteristic meteoroid dimensions.

Moreover, a single collision of an atmospheric molecule with the surface of a meteoroid may eject up to 500 meteoric atoms and molecules (Jenniskens et al. 2000); some which attain axial velocities 1.5 times higher than the parent meteoroid (Rajchl 1969). The hydrodynamic shielding becomes effective when the mean free path within the vapour cloud is approximately an order of magnitude smaller than the radius of the meteoroid (Popova et al., 2003). The meteor shock wave is formed when the hydrodynamic shielding is compressed (at lower altitudes), such that the changes in velocity, temperature and density are essentially a discontinuity.

The observational evidence indeed shows that the meteor bow shock (initial shock envelope associated with hypersonic flows) and the cylindrical shock wave (essentially approximated as a blast wave from the line source which depends on the amount of energy deposited per unit length (Lin 1954)) appear much earlier than predicted by classical gas dynamics theory (e.g. Rajchl 1972; Bronshten 1983; Brown et al. 2007; Silber & Brown 2014). This occurs before the onset of the continuum flow (e.g. Probstein 1961; Bronshten 1983) and for most meteoroids takes place in the lower region of the transitional flow regime. This is especially relevant for overdense meteors discussed in our study. We expand this topic in more detail in the sections to follow.

The atmospheric gases swept behind either the hydrodynamic cap or, at lower altitudes, overdense meteor shock wave front (Figures 1 and S1) are dissociated and ionized. The high energy inelastic collisions of atoms or molecules behind the shock front and in the flow field, result in a change of internal state and velocity of atomic and molecular species (Schunk & Nagy 2009). Furthermore, these collisions usually involve the exchange of translational, rotational, and vibrational energy, leading to the subsequent formation of new species (i.e. Brun 2009; Berezhnoy & Borovička 2010). During elastic collisions in the overdense meteor wake, which occur in the 'lower' energy regime, the momentum and kinetic energy of the colliding particles are conserved and only translational energy exchange takes place. We are however not concerned with the processes that occur much farther back in the meteor wake and within the initially formed volume of the meteor train. This is because these processes do not contribute to appreciable removal of electrons from the overdense meteor train. Consequently, in this work we are mainly interested in the processes that occur in the expanding high temperature meteor train boundary with the ambient atmosphere. The processes within the wake and immediate train of the meteor trail have been discussed by Menees & Park (1976), Park & Menees (1978), and Berezhnoy & Borovička (2010).

In the initial phase of the overdense meteor trail evolution, the ablated meteor plasma radiative energy loss takes place during the collisional deceleration, where the ablated plasma and the initially entrained and modified ambient gas stop within several hundred meters (Jenniskens et al. 2004). These processes coincide with what we define in this paper as the first stage of the overdense meteor trail evolution. This dynamic evolution of the high temperature ablated plasma and vapour exhibits rapid and highly turbulent initial flow in the meteor wake, which leads to the adiabatic formation of the more dynamically stable meteor trail volume with the initial radius $r_0$ (e.g. see Lees and Hromas 1962; Jones 1995; Jones and Campbell-Brown 2005; Hocking et al. 2016).

As shown in Figure 1, the formation of the initial meteor trail is preceded by the cylindrical shock wave (the latter depends on the pressure ratio behind the shock and the ambient atmosphere, as a function of meteoroid size, velocity, ablation rate and Knudsen number; these are discussed in detail in Sections 3 and 4 and the Supplementary Material). The cylindrical shock wave rapidly merges with the bow shock wave and expands radially with velocities significantly lower than the entry velocity of the meteor (in the considered meteor velocity range, e.g. see Tsikulin (1970)). However, the cylindrical shock wave (discussed in detail further in the text) is sufficiently strong that it results in a near-instantaneous rise in temperature immediately behind the shock front which is of the order of several thousand kelvin.

The second stage of high temperature overdense meteor trail evolution is characterized by onset of ambipolar diffusion (Francey 1963; Pickering & Windle 1970) which takes effect immediately after the explosive formation of the initial meteor trail volume (Figure 1). It should be noted that the rate of ambipolar diffusion is a function of temperature and pressure (Hocking, Thayaparan & Jones 1997).

The initial exchange and equilibration of translational, rotational, and vibrational energy between atmospheric and ablating meteor constituents trapped within the flow-field brings the temperature in the wake of the meteor train (and initially formed meteor trail volume with radius $r_0$) down to about 4,400 K (Jenniskens et al. 2004). Additionally, in the aforementioned important study, Jenniskens et al. (2004) found a marginal rise in temperature with decreasing altitude. More importantly though, they observed comparatively constant temperatures in the velocity range between 35 km/s and 72 km/s and masses between $10^{-5}$ g and 1 g. The lower end of the spectrum of mass values reported by Jenniskens et al. (2004) is consistent with strong underdense meteors, while the upper end of the reported values corresponds to overdense meteor parameters.

Furthermore, the authors established that faster and more massive meteoroids produce larger emission volume, but not a significantly higher air plasma temperature. Comparing their data with fireball temperatures obtained earlier, Jenniskens et al. (2004) concluded that the variation of meteor plasma emission temperatures for meteoroids in the range of masses between $10^{-5}$ g and $10^6$ g is only up to several hundred kelvin. Indeed, while surprising, such behaviour can be easily understood in terms of energy loss to molecular ionization, dissociation and hyperthermal chemical reactions in hypersonic reactive flows (Zel'dovich & Raizer 2002; Anderson 2006; Brun 2009).

Moreover, observations show that it takes a few seconds for the temperature in a fireball wake with a visual magnitude of -12 to cool down from 4,500 K down to 1,200 K, while for a typical overdense meteor with $M_v$ = -3, it takes ~0.1 s (Jenniskens 2004). This reported meteor train cooling time is a significant development in understanding of the early meteor trail evolution, along with the observed and reported temperature values, because it allows a substantial amount of time for large scale hyperthermal chemistry to take place on the boundary of the expanding overdense meteor train. More energetic and perhaps more complex sets of hyperthermal chemical reactions which occur inside the meteor train were discussed by Menees and Park (1976), Park and Menees (1978), and Berezhnoy and Borovička (2010). The implications of this will be discussed further shortly.

**2.2 Hyperthermal Chemistry Within the Trail**

We now turn to issues of chemistry. While the chemistry within the trail is fairly well understood for the case that the trail has cooled down to ambient atmospheric temperatures (referred to as "thermalized chemistry", e.g. Baggaley 1978; 1979; Plane 2012; Plane et al. 2015), there may also be substantial chemistry in the early stages of the trail formation, when temperatures are still very hot. This has not been explored as thoroughly as the thermalized chemistry, but has the potential to have significant impact on the life-cycle of the trail. This chemistry can occur in various places, including in association with the shocks (e.g. Zel'dovich & Raiser 2002), within the trail, and (notably) on the edge of the trail. These processes can potentially result in rapid electron removal from the boundaries of the postadiabaticaly expanding high temperature meteor train. Eventually, this short period of initial electron removal terminates relatively rapidly, and then ambipolar diffusion takes over until the time at which thermalized chemistry starts to play the dominant role of electron removal (Baggaley & Cummack 1974; Baggaley 1978). However, the time taken for the hot parts of the trail to settle down to ambient temperature is still the subject of considerable uncertainty, and can have profound effects on diffusion rates. Hocking et al. (2016), appendix C, suggests that this may require that the net diffusion rates of the trail should be the geometric average of the diffusion coefficients of the hot plasma and the ambient background atmosphere. The validity of this assumption is critically dependent on the rate at which temperature equilibrium in the trail is achieved. While once considered near-instantaneous, this is now questionable. We will return to this point later: for now, we concentrate on the chemical processes that occur while the region is still hot, which we take to be prior the first 0.1 to 0.3 seconds, which is required for the overdense trail to thermalize.

We now look at the implications for chemistry within the high temperature regime. This high temperature meteor train expands post adiabatically into the ambient atmosphere, modified by the cylindrical shock wave.

These processes then enable the temperature driven oxidation of meteoric metal ions in the trail boundary by the ambient oxygen that survives the passage of cylindrical shock wave some distance away from the high temperature meteor train and also the shock dissociated product ($O_2$) of ozone ($O_3$) in the meteor near-field. The reaction is expressed as $M^+ + O_2 \rightarrow MO^+ + O$, where $M^+$ is a common meteoric metal ion. The process is generally completed in $10^{-3} – 10^{-1}$ s, for altitudes between 80 and 100 km. The observational evidence of much slower thermalization of the meteor trains (Jenniskens et al. 2004) corroborates the presence of a high temperature environment conducive to hyperthermal chemistry. Notably, the production of metal oxide ions will be governed by the second Damköhler number (which represents the ratio of the chemical reaction rate to the ambipolar diffusion mass transfer rate) and the temperature (1,500 K < T < 3,000 K), with the highest yield at about 2,500 K (Berezhnoy & Borovička 2010). We will discuss this in detail in the next section.

The third and final stage of the trail development takes place within the boundary of the almost thermalized ambipolarly diffusing meteor train sketched in Figure 1. Here, hyperthermally-produced meteor metal oxide ions rapidly remove electrons in the almost thermalized train, through temperature independent dissociative recombination, $MO^+ + e \rightarrow M + O$ (Plane 2012; Plane et al. 2015). The reaction terminates when $MO^+$ is consumed. Depending on the available raw material ($MO^+$), this reaction may have a noticeable impact on the lifetime of the trail by removing electrons in this early phase.

In the following sections we examine in more detail the evidence for hyperthermal chemistry which has so far been only briefly summarized. We also refer to the modelling work used to illustrate the effect of shock waves; details are included in the Supplementary Material, where we present an even more comprehensive discussion.

## 3. Linking the Shock Waves and Meteor Train – Atmosphere Hyperthermal Chemistry

The meteor cylindrical shock waves have the strongest effect in the region of the ambient atmosphere relatively close to the adiabatically formed meteor trail volume. In this region, defined as the characteristic or blast radius $R_0$ (ReVelle 1976; Silber, Brown & Krzeminski 2015), the initial energy deposition per unit length is the largest, because overdense meteor cylindrical shock waves are in principle approximated as explosive line sources (e.g. see Lin 1954; Bennett 1958; Jones et al. 1968; Tsikulin 1970). That is primarily due to the fact that there is almost an instantaneous release of comparatively large quantity of energy in a limited geometrically defined space (Steiner & Gretler 1994). We note that $R_0$ (the characteristic or blast radius) is different than the previously defined $r_0$ (meteor trail volume radius). The relationship between the maximum energy deposition and the characteristic radius $R_0$ shown in Figure 2 (ReVelle 1974, 1976), is expressed as:

$R_0 = (E_0/p_0)^{0.5}$ (1).

Here, $E_0$ is the energy deposited per unit path length (which in the case of a meteoroid is the same as the total aerodynamic drag per unit length) and $p_0$ is the ambient pressure (e.g. Silber et al. 2015). The term characteristic radius is used only in reference to strong shock waves, when the energy release ($E_0$) is sufficiently large that the internal energy of the ambient atmosphere is negligible (Lin 1954; Hutchens 1995). Figure 2 shows the initial radius ($r_0$) of bright overdense meteors (Baggaley & Fisher, 1980; Ceplecha et al. 1998) and the radius of the overdense meteor trail after 0.3 s. Those are compared with the characteristic radius ($R_0$) associated with the constant energy deposition of 100 J/m and 1000 J/m, for the altitude 80 – 100 km. The aforementioned energies represent the velocity, size and composition dependent peak energy depositions (e.g. see Zinn, O'Dean & ReVelle 2004; Silber et al. 2015) for different sizes of overdense meteors ablating in that narrow region of MLT. It is readily seen that $R_0$ is always greater than $r_0$ for constant energy deposition values corresponding to overdense meteors in MLT and it approximately matches or is greater than the radius of the ambipolarly expanded meteor trail volume after 0.3 s (depending on the choice of initial $r_0$ and the diffusion coefficient $D$). We will come back to this point later (Section 4.1).

However, the maximum effect on the ambient atmosphere, such as dissociation, is most dominant within the characteristic radius of $R_0$. After that, the shock wave attenuates rapidly and transitions to the acoustic regime within $10R_0$ (ReVelle 1976; Silber 2014; Silber et al. 2015). The initial temperature behind an overdense meteor cylindrical shock wave is typically in the vicinity of 6,000 K (as it will be demonstrated in the following sections), sufficient to dissociate $O_2$ and $O_3$ within $R_0$.

Oxygen, regardless of its initial source (Dressler 2001), is the most likely molecule to react hyperthermally and rapidly with the ablated meteor ion in the boundary of the high temperature meteor train (e.g. see Murad 1978). For the simplicity of the exposition, this paper consequently only focuses on the initial meteor train near-field (~$R_0$) where the product of high temperature oxidation of meteor metal ions, along with subsequent dissociative recombination, is the only reasonably fast mechanism capable of removing electrons from the boundary of meteor trail in the initial stage of postadiabatic ambipolar diffusion (Dressler 2001). Another important aspect examined in this work is the source of $O_2$ (ambient or products of ozone shock dissociation) which dominates in the high temperature rapid production of the meteor metal ions oxides that are subsequently responsible for the posthyperhermal chemical removal of electrons from the boundary of the overdense meteor train.

To further examine these issues, we need to consider the pressure and temperature gradients in the flow field in and around the meteoroid and in the meteor wake relative to the ambient atmosphere, along with the dissociative behaviour, excitation and ionization potentials of atmospheric molecules. Furthermore, some aspects of high temperature gas dynamics and chemistry involving both major and minor MLT species need to be further illuminated in order to understand the complex processes that take place on short time scales in the boundary of the postadiabatically expanding hot meteor train.

### 3.1 Initial Shock and Hyperthermal Chemistry Whithin High Temperature Meteor Trail

Within the volume of vapour and plasma, beginning from the region behind the meteor shock front and enclosed by the envelope of the initial shock (see Figure S1), various complex physico-chemical processes, such as the high-temperature reactive and non-equilibrium flows, ionization, dissociation and excitation, take place at very small time scales (Park & Menes 1978; Menees & Park 1976; Kogan 1969; Shen 2006; Brun 2009, 2012). The relative importance and the rates of those processes depend upon the temperature and density and the time scales at which the relaxation between translational, rotational and vibrational energies take place.

The early chemical reactions occur as a result of almost instantaneous gas heating, which is caused by collisions with ablated and evaporated meteoric material. These collisions, however, are caused by initial "instantaneous" compression and kinetic and radiative energy exchange behind the shock front (Anderson 2006). High energy molecular and atomic collisions also occur (at temperatures generally far above the characteristic vibrational temperature of the diatomic molecule), followed by the equilibration between translational and internal degrees of freedom. This is superseded by subsequent dissociation, ionization and radiation in addition to various non-equilibrium chemical reactions (Panesi et al. 2011; Brun 2012).

Behind the initial meteor shock-front (Figure S1), velocity dependent ionization occurs rapidly, involving both impinging atmospheric constituents and ablated meteoric atoms. Notably, meteoric metals atoms (e.g. Fe, Mg) will ionize more efficiently due to their lower ionization potential (Dressler 2001). Competing ionization processes take place, such as ionization by molecular and atomic collisions, electron impact and ion impact ionization, in addition to photo ionization; the respective reactions and required energies are given and discussed by Lin & Teare (1963); Park (1989); Starik, Titova & Arsentiev (2009). Moreover, behind the shock wave front, the vibrational temperature depends on vibrational relaxation rates, as well as coupling of the vibrational relaxation and dissociation of molecules (Zabelinskii et al. 2012). However, the rate of dissociation behind the shock wave is reduced when the vibrational temperature has not equilibrated with the translational temperature (Boyd, Candler & Levin 1995).

The translational temperature, which increases rapidly behind the shock front (e.g. Sarma 2000; Boyd 2000; Zinn et al. 2004; Zinn & Drummond, 2005), decreases quickly as the rotational and vibrational energies are raised. The vibrational modes take longer to equilibrate with translational and rotational temperatures. In the case of the reactive flow around overdense meteors in the MLT region, the comparison of chemical and hydrodynamic timescales during the initial stages of the flow within the shock layer, as depicted in Figure S1, indicates that the equilibrium is still not reached in the initial flow field behind the shock front because the chemical reaction time scales are longer than the hydrodynamic timescale (Berezhnoy & Borovička 2010). However, after equilibrium between the various energy modes is established, further energy is consumed by dissociation and ionization (Hurle 1967), followed by the beginning of various thermally driven chemical reactions with different characteristic times (Sarma 2000; Brun 2012).

Further down the meteor axis, within the high temperature region in the meteor wake, shock-modified reactive flow of ablated vapour and plasma occurs, carrying the entrained excited, dissociated and ionized atmospheric constituents. This is an ideal environment for the formation of nitric oxides, as was discussed by Menees & Park (1976) and Park & Menees (1978).

However, it is useful to recall at this moment that the dissociation and ionization threshold energies of $N_2$ and $O_2$ are very high (Massey & Bates, 1982; Rees 1989). Comparatively, meteoric metal atoms have low ionization potentials and can be ionized efficiently, as mentioned earlier (relative to atmospheric molecules and atoms), in high velocity neutral collisions (Dressler 2001).

It must be emphasized at this point that in general, no appreciable electron removing reactions between the meteoric constituents take place within the expanding meteor train (Berezhnoy & Borovička 2010). This is important as it indicates that the processes responsible for the initial rapid and short lasting electron removal occur mainly on the boundary of the meteor train.

Within the high temperature meteor trail, nitric oxide is generally formed by a hyperthermal reaction between available $N_2$ and O within the meteor trail volume with the initial radius $r_0$, where $N_2 + O \rightarrow NO + N$. The reaction proceeds when the temperature is in the range between 2,000 K and 10,000 K (Menees & Park 1976). Below 2,000 K, NO is further produced by the reaction $N + O_2 \rightarrow NO + O$. The first reaction is endothermic, while the second reaction is temperature independent, and will proceed inside the volume of the adiabatically formed meteor train with the initial radius $r_0$ until almost all supplies of N atoms are exhausted (including small quantites of $O_2$ within the high temperature meteor train). It should be noted that the reverse of the first reaction occurs at lower temperatures, which removes N and NO from the flow (Menees & Park 1976).

From the perspective of the high temperature meteor train chemistry, this is very relevant, because there will be a negligible amount of remaining N within the meteor train volume to engage in reactions outside the meteor train boundary.

A more detailed study and description of the thermally driven chemical reactions in high temperature meteor train is given by Berezhnoy & Borovička (2010). A detailed analysis of the reactions of atomic and molecular metastable species behind the shock wave is presented by Starik et al. (2009). The authors presented an extensive list of reactions and reaction rates for the range of excited dissociated and ionized atmospheric constituents, which serves to further illuminate the very complex and previously difficult-to-model chemical dynamics of the shock wave environment. We can now go back and examine the dynamics and physico-chemical effects of the meteor cylindrical shock waves.

## 4. Evaluation of Shock Wave Effects

### 4.1 Dynamic and Physico-Chemical Effects of Overdense Meteor Cylindrical Shock Waves

Depending on Knudsen number, velocity, size and composition of overdense meteors, the energy deposited per unit path length may reach as high as several thousand J/m, (Zinn et al. 2004; Silber et al. 2015). This energy, assumed to be released instantaneously along the axis of meteor propagation, drives the radial expansion of the cylindrical shock (Zel'dovich & Raizer 2002). In treatment of the cylindrical shock waves, it is assumed that all of that energy is deposited almost instantaneously in the cylindrical volume of the atmosphere with radius $R_0$ (Lin 1954; Plooster 1968; Tsikulin 1970), as mentioned earlier.

It is well established that the speed of the shock wave depends only on the difference in pressure of the region where the energy is deposited relative to the pressure in the ambient gas (e.g. Hurle 1967). Thus if the velocity (and consequently the strength) of the meteor bow and vapour cylindrical shock waves are to be determined, the pressure behind the shock front or vapour pressures in the compressed flow field region behind the meteoroid are important parameters and must be known (Bronshten 1983; Zel'dovich & Raizer 2002; Anderson 2006). While we can distinguish, for pedantic purposes, between the two main types of the cylindrical shock waves (the initial bow shock and the ablation amplified recompression cylindrical shock wave; see Supplementary Material) during the initial shock evolution (e.g. Hayes & Probstein 1959;

Bronshten 1983; Sarma 2000), that distinction cannot be made outside of the immediate region of maximum energy deposition with the characteristic radius $R_0$, as these two types of shock waves will coalesce rapidly. The pressures in the stagnation region behind the region of the blunt shock front ahead of and on the axis of the meteor can be determined based on the meteoroid characteristics (Bronshten 1965; Tsikulin 1970).

Let us consider and compare the bow (initial or primary cylindrical shock wave) and the ablation amplified recompression cylindrical shock wave (Sarma 2000) which, in the case of ablating meteoroids, is defined as the ablational or vapour cylindrical shock wave (Bronshten 1983). In simple terms, the bow shock wave strength and the velocity of radial expansion will depend on the meteoroid velocity, the initial flow translational temperatures, and the subsequent pressures behind the shock front in the front of the meteor. It will also strongly depend on the specific heat ratios, as they will dictate the geometry of the blunt region (Anderson 2006). The recompression shock wave, while common in all hypersonic bodies (e.g. Hayes & Probstein, 1959; Sarma 2000), will be different for overdense meteors in the transitional and continuum flow regimes, because it will depend directly on the amount of ablated material from the meteoroid (Bronshten 1983; Zinn et al. 2004). As importantly, the strength of the cylindrical shock originates from the compressed flow field around and behind the meteoroid (e.g. the neck region of the flow field), and depends on both the flow temperature, and vapour and plasma pressure at the neck (the region of the maximum gas and plasma compression behind the meteoroid) (Figure S1).

For illustrative purpose, consider the dissociated ambient atmosphere, initially swept behind the meteor shock front. It is compressed along with ablated meteoric plasma and vapour (Popova et al. 2001) and still has temperature significantly greater than 10,000 K in the immediate flow field behind the meteor (Boyd 2000; Jenniskens et al. 2000) (Figure S1). In that region, the flow field converges and is compressed to pressures several orders of magnitudes higher than the ambient atmospheric gas. Moreover, the pressure increase relative to the ambient gas is amplified by ablation, which frees significantly more than $10^{16}$ ions and atoms per meter, for the case of an average overdense meteor. Considering that the ambient gas temperature is about 200 K, and the temperature in the flow field behind the meteoroid exceeds $10^4$ K, it can be shown using the equation of state for a gas that even without ablation, or a volume reduction, the pressure increase in the flow field behind the meteoroid exceeds 50 times that of the ambient gas.

This problem was first considered by Dobrovol'skii (1952), and while initially dismissed by relevant investigators at the time, it has been proven valid (Bronshten 1983). Let us consider the loss of meteoroid kinetic energy, which can be written using the following expression:

$$\frac{dE}{dt} = \frac{d}{dt}\left(\frac{mv^2}{2}\right) = mv\frac{dv}{dt} + \frac{v^2}{2}\frac{dm}{dt} \qquad (2).$$

Here, the first term on the left represents the energy lost per unit of time and *m* and *v* are the meteoroid mass and velocity, respectively (e.g. Romig 1964; Gritsevich & Koschny, 2011). Dividing both sides of (2) by the velocity (*v*) (Bronshten 1983) the energy deposition per unit path length can be obtained:

$$\frac{dE}{dl} = m\frac{dv}{dt} + \frac{v}{2}\frac{dm}{dt} \qquad (3).$$

As indicated by Bronshten (1983), the first term on the right in equation (3) is the energy used to form the bow shock wave, assuming no ablation. The second term then is the energy partitioned to the ablation and lost to the ablated vapour per unit length. It can be shown that the second term is utilized to describe the formation of the comparatively stronger cylindrical vapour shock wave. To demonstrate this, we need to consider the ratio of the differentiated terms on the RHS in equation (3). The ratio of the two terms on the right in equation (3), $\frac{v}{2}\frac{dm}{dt} / m\frac{dv}{dt}$ were compared by Dobrovol'skii (1952), with certain simplifying assumptions (see Bronshten (1983) for a discussion), and the results indicated that the second term is significantly bigger. Depending on the meteoroid velocity and the rate of ablation, the second term may be more than two orders of magnitude larger than the first term, especially for the higher velocities.

As discussed earlier, this meteoroid-deposited energy can be equated with the blast wave from exploding cylindrical line sources as discussed by Lin (1954), Bennett (1958), Jones, Goyer & Plooster (1968), and Plooster (1970). Lin (1954) presented the solution for the cylindrical shock wave produced by instantaneous energy release, where he defined the radius of the cylindrical shock wave and determined its rapid decay as a function of time. The shock envelope behind the meteoroid is a function of the aerodynamic drag, initial density and meteoroid velocity. Bronshten (1983) offers a detailed meteorcentric discussion of the problem.

The pressure ratios of the ablated, vapourized meteoroid and plasma in the flow field ($p$) to that of the ambient atmosphere ($p_0$) for an average size of overdense meteors with 1 cm radius, as evaluated by Bronshten (1983) lie in the range $10^2 < p/p_0 < 10^4$. This is particularly true for events with velocities exceeding 30 km/s, where much more energy is transferred to the flow field vapour and plasma behind the shock front, than is spent on the ablation process. It is the dispersion of this ablated and pressurized "vapour" in the front of the meteoroid that amplifies the shock wave (Dobrovol'skii 1952; Bronshten 1983; Zinn et al. 2004).

Consequently, it can be seen that the ablation and vapour amplified cylindrical shock wave is, in the case of intensely ablating, fast meteoroids, significantly stronger than the initial bow shock wave in the absence of ablation. In principle however, the two cylindrical shock waves (initial bow, and ablation amplified recompression or cylindrical shock wave) rapidly merge and cannot be distinguished, as mentioned earlier.

We can use this value of $p/p_0$ to estimate the strength of the typical overdense meteor cylindrical shock wave, assuming that we know- the pressures and temperatures of the vapourized and ablated material, as well as the flow field surrounding the entrapped atmospheric dissociated molecules, in the neck region behind the meteoroid (Figure S1).

For the purpose of simplification, we assume that the pressures of the ablated high temperature vapour and plasma around and behind the meteoroid exceed the ambient pressure by at least two orders of magnitude (Bronshten 1983). In principle, this might be a significantly understated value as demonstrated in early studies (Bronshten 1965, 1983) and it can be reasonably interpreted to correspond more to the pressure ratios associated with transitional meteors (e.g. Popova et al. 2000; 2001). However, we use it here for expository nature of the problem. Furthermore, it is reasonable to assume that the pressure in the neck region of the flow field, behind the meteoroid, due to the ablation, dissociation and high temperature flows will be similar to pressures in the region around the stagnation point (Figure S1), behind the initial shock front. Therefore, this gives us a reasonable tool to approximately evaluate the cylindrical shock velocity and strength that originates from the high temperature compressed flows in the neck region (Figure S1).

Thus, in order to evaluate the initial velocity and strength, and consequently the effects of the cylindrical shock waves, we need to know with reasonable accuracy the vapour and plasma pressure behind the cylindrical shock front in the flow field region behind the meteoroid where the shock wave is generated.

For the purpose of this exposition, we consider Brohnsten's (1983) discussion as a guide to approximate the pressure ratios of ablated vapour and the ambient atmosphere as $p/p_0 \sim 100$ as discussed above. This is taken as the lowest value for overdense meteors with a size range discussed earlier for the purpose of evaluating the strength of the cylindrical shock wave (see Chapter 3, Section 17 in Bronshten 1983 for discussion). The cylindrical shock wave velocity (or Mach number) can be easily obtained from the expression for the pressure behind the shock front which is generally evaluated using the Hugoniot relationship. It relates the vapour pressure behind the shock ($p$) and the ambient pressure ($p_0$), shock Mach number ($M_{sw}$) and the ratio of specific heats ($\gamma$) (e.g. Lin 1954; Jones et al. 1968; Tsikulin 1970):

$$\frac{p}{p_0} = \frac{2\gamma}{\gamma+1} M_{sw}^2 \qquad (4).$$

This relationship can be used in the region of the strong shock wave where $p \gg p_0$ (Lin 1954; Jones et al. 1968). Another way to roughly estimate the shock wave velocity is by utilizing the vapour temperature and corresponding high thermal velocities, as demonstrated by Zinn et al. (2004).

Experimental observations agree with expression (4) in the region of the strong shock wave ($R_0$) (e.g. Jones et al. 1968, Plooster 1970 and references therein) where empirically derived relations for the density, pressure and temperature ratios (see Zel'dovich & Raizer, 2002) are written as:

$$\frac{\rho}{\rho_0} = \frac{6}{1+5M_{sw}^{-2}} \qquad (5)$$

$$\frac{p}{p_0} = \frac{7}{6} M_{sw}^2 - \frac{1}{6} \qquad (6)$$

$$\frac{T}{T_0} = \frac{1}{36}(7 - M_{sw}^{-2})(M_{sw}^2 + 5) \qquad (7).$$

Here, $\rho_0$ and $T_0$ are the mean density and temperature ahead of the shock wave, respectively. The same parameters without subscript are the values just behind the shock front. Combining equations (5-7) (for details see Zel'dovich & Raizer, 2002; Hurle 1967), the temperature behind the shock wave can be calculated if the cylindrical shock wave Mach number or the pressure ratio are known. The discussion and theoretical treatment of shock wave Mach number along with the flow regimes are given in Section S1 (Supplementary Material) in this paper.

Then, using the Rankine-Hugoniot relations ($p/p_0 \approx 100$), and assuming an ideal diatomic gas, the velocity of the cylindrical shock wave is calculated to be around Mach 9.3 and the temperature behind the shock is in the range of 6,000 K, assuming an ideal gas (see Hurle (1967), where the value of 6,020 K is suggested).

However, the actual value of temperature is lower, as the non-ideal gas behaviour affects the temperature values through the mechanisms (Anderson 2006) discussed earlier in the text and in considerable detail in the Supplementary Material. Moreover, in the MLT, the ratio of specific heats ($\gamma$) will be also different, leading to the lower values of the calculated temperature (e.g. Viviani & Pezzella, 2015). Generally, below 95 km, the pressures of the vapour and the rate of ablation for average overdense meteors will be in the above mentioned range, depending on velocity (Bronshten 1983; Boyd 2000). However, considering the much higher vapour pressure ratios for typical chondritic meteoroids behind the shock front, the velocity of the ablationally amplified cylindrical shock wave (within the $R_0$ region) may be significantly higher than our estimate. In reality, the Mach number of the cylindrical chock waves may easily reach or exceed Mach 20 for the upper sizes of overdense meteors with higher velocity and large energy deposition ($E_0 > 1000$ J/m).

These high velocities bring the temperatures behind the cylindrical shock wave to the range of 6,000 K, even after taking into consideration a non-ideal gas specific heat ratio, and effects of relaxation and dissociation (Anderson 2006).

Importantly, such temperatures behind the cylindrical shock wave are sufficient for strong dissociation and excitation of atmospheric species within $R_0$, but will not be high enough for any appreciable ionization. Generally, at temperatures in the range 3,000 - 7,000 K behind the shock front in a typical atmospheric diatomic gas, there is still no appreciable ionization. Under such conditions, the molecular vibrations are excited relatively quickly, and the thickness of the wave front is connected with the slowest relaxation process, namely molecular dissociation (Zel'dovich & Raizer 2002).

Knowledge about these temperatures is the key to understanding the dominant chemistry in this region. Let us consider molecular oxygen first, which is inert below 800 K (Zel'dovich & Raizer 2002). The dissociation energy for $O_2$ is 5.12 eV or about 59,000 K (Bauer 1990). The rate of dissociation of oxygen behind the shock wave is a function of temperature, as shown by Ibraguimova et al. (2012). The heated $O_2$ molecules begin to dissociate between approximately 2,000 K and 4,000 K at normal pressure, while above 4,000 K almost all oxygen is dissociated (Bauer 1990; Anderson 2006). However, the dissociation temperature range is affected by the pressure, and thus in the MLT region, the dissociation takes place at lower temperatures (Bauer 1990). The dissociation of molecular oxygen in the atmosphere has been presented in detail by Nicolet & Mange (1954) and will take place behind the strong shock wave, when the vibrational temperature is equilibrated with translational and rotational temperatures. As a comparison, the dissociation energy of $N_2$ is 9.76 eV (Bauer 1990). $N_2$ dissociation starts slowly above 4,000 K and is almost complete just above 9,000 K (Anderson 2006; Fridman 2008).

Moreover, the thermal non-equilibrium chemical kinetics and dissociation of $O_2$ behind the shock front has been investigated by Gidaspov, Losev & Severina (2010). For typical cylindrical shock wave strengths discussed here, the time scales for the dissociation of $O_2$ will be closely in line with the typical time of vibrational excitation of two-atom molecules, and can be approximated by the Landau-Teller formula (Gidaspov et al. 2010 and references therein). Both vibration relaxation and dissociation time scales decrease with increasing temperature, and their ratio for oxygen approaches unity in the region between 2,000 K and 4,000 K. For the purpose of this analysis, and based on the data and calculations presented in Nelson (1964), Bauer (1990) and Takayama (2012), and the rates given by Gidaspov et al. (2010) and Ibraguimova et al. (2012), it is reasonable to estimate that a significant proportion of $O_2$ is dissociated within approximately $10^{-4}$ s behind the cylindrical shock wave passage within the $R_0$ blast region surrounding the initially formed meteor trail volume in the MLT region.

The implication of the above discussion is that behind the overdense meteor cylindrical shock wave in the near field region surrounding the meteor train boundary, and at related temperatures, most $O_2$ will be dissociated in the proximity of the boundary of the initial meteor high temperature train volume and less so toward the $R_0$, with only a negligible fraction of $N_2$ dissociated. However, it must be emphasized again, that while we are discussing the generalized case, the actual amount of surviving $O_2$ depends primarily on the energy deposition which may vary across the overdense meteor size spectrum.

For comparative purposes and in order to investigate the flow and temperature fields around a meteor, we have numerically modeled the hypersonic flight dynamics for two non-ablating spherical bodies (diameter = $2.5 \cdot 10^{-2}$ and $1 \cdot 10^{-1}$ m) with velocity 35 km/s and at 80 km altitude (Section 4.3).

**4.2 Ozone Dissociation Behind the Cylindrical Shock Wave**

Ozone in its native form cannot survive the effects of the meteor generated cylindrical shock waves because of its characteristic temperature sensitivity (Schumacher 1960; Jones & Davidson 1962; Benson & Axworthy 1965; Michael 1971) and its low dissociation potential (Center & Kung 1975; Endo et al. 1979), which is much lower than that for $O_2$. However, as a result of such properties of ozone, a variety of energetic species can be produced through its dissociation, with modest amounts of input energy (Zel'dovich & Raizer 2002). Let us consider what happens with ozone at those initial kinetic temperatures in the range of 6,000 K behind the cylindrical shock front, in the meteor trail near-field. The process of collisional dissociation of ozone with sufficiently energetic particles in the high temperature region behind the shock wave usually corresponds to the reaction:

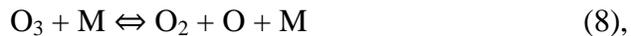
$O_3 + M \Leftrightarrow O_2 + O + M$ (8),

where M is any atmospheric molecule or atom ($O_2$, $N_2$, O, N). Ozone dissociation (Jones & Davidson 1962; Center & Kung 1975; Fridman 2008; Konnov 2012) is highly temperature dependent, and at meteor shock wave temperatures in the MLT region, the dissociation times are on order of $10^{-5}$ s. The ozone dissociation time decreases with increasing temperature $T_0$. It should be also noted that while the dissociation of ozone is an endothermic reaction, the formation of ozone is exothermic. However, the higher temperatures in the vicinity of the boundary region of the meteor train impede new ozone production in the shock heated gas and meteor plasma (from the initially dissociated oxygen and ozone products). The kinetics of the excited products of dissociation of $O_3$ and $O_2$ is discussed by Yankovsky & Manuilova (2006).

Comparing the dissociation potentials of oxygen and ozone, we can see that relative to oxygen dissociation energy of 5.12 eV, the dissociation energy of ozone is significantly less, at about 1.04 eV (Bauer 1990).

Above 1,500 K, ozone dissociation times are in the order of μs (Jones & Davidson 1962; Benson & Axworthy, 1965; Johnston 1968; Michael 1971; Center & Kung 1975; Konnov 2012; Peukert et al. 2013). Endo et al. (1979) investigated the thermal dissociation of $O_3$ behind a shock wave with temperatures between 600 K and 1,100 K, where ozone dissociates into the low energy triplet $O_2(X^3\Sigma_g^-)$ and the low energy molecular oxygen $O(^3P)$ with activation energy of 0.98 eV.

The decomposition of ozone in flames, recently modeled by Konnov (2012), is known to produce the triplet $O_2(X^3\Sigma_g^-)$ and $O(^3P)$, and yields small quantities of the singlet $O_2(a^1\Delta_g)$. At shock wave temperatures and time scales, the reaction: $O_2(a^1\Delta_g) + O_2 \rightarrow O_3 + O$, along with other ozone-forming reactions, will not proceed. Moreover, the collisional efficiency of $O_3$ is assumed to be 2.5 – 3 times higher than that for $O_2$ (which is in turn more collisionally efficient than $N_2$). Notably, atomic oxygen has a collisional efficiency that is 4 – 5 times more than that of $O_2$ (Makarov & Shatalov 1994; Luther et al. 2005; Konnov 2012). However, considering the meteor cylindrical shock wave temperatures and the associated short time scales, a significant amount of $O_2$ that originates from the $O_3$ shock dissociation will survive closer to the meteor trail boundaries (within *R₀*). This is due mostly to the finite energy budget and finite timescales available for dissociation, which exist behind the typical overdense meteor cylindrical shock waves. Of course, the same consideration can be applied to fireball-type of events (as demonstrated by Zinn et al. (2004)) or on the opposite end of the size-spectrum, to transitional meteors, where the strength of the cylindrical shock remains uncertain (in cases when such shocks exists). As will be demonstrated shortly, this is very important, as surviving oxygen is available for the high temperature reactions with meteor metallic ions and is of critical importance for determining any potential role of these processes in early electron removal.

At overdense meteor cylindrical shock wave temperatures, however, dissociation of ozone will yield the presence of both excited and ground state particles of both $O_2$ and O, where the excited species $O_2(a^1\Delta_g)$ and $O(^1D)$ will be present in relatively small quantities (Park 1989; Klopovskii et al. 1995; Starik et al. 2010). However, the ground state species are the primary product of ozone shock dissociation. In terms of excited species, the resulting singlet O($^1$D) is rapidly quenched (collisionally de-excited) by collisions with the ambient molecules, atoms and electrons (e.g. $N_2$, $O_2$, O) and subsequently consumed by $N_2$ (Zipf 1969; Capitelli et al. 2000; Fridman 2008; Schunk & Nagy 2009).

Metastable $O_2$ ($a^1\Delta_g$) is relatively immune to quenching by a major atmospheric gas (Zipf 1969), and may react under favourable high temperatures with meteoric metallic ion such as $Fe^+$ and $Mg^+$.

Consequently, the thermally driven reactions between meteor metal ion species $M^+$ and oxygen (remaining from shock dissociation of $O_3$ and ambient $O_2$ that survived the shock wave) will proceed in the boundary region of the hot meteor trail as long as there are favourable temperature regimes:

$$M^+ + O_2 \rightarrow MO^+ + O \qquad (9)$$

However, the Maxwell-Boltzmann distribution of shock modified species behind the shock wave (Cercignani 2000), indicates that the surviving quantities of the ambient molecular oxygen will also participate in the same thermally controlled reaction, albeit toward the outer boundary of the $R_0$ region (our modelling results confirm this, see Section 4.3). However, in terms of the overall contribution to the reaction (9), oxygen from shock dissociated ozone may not play a dominant role within $R_0$, because the $O_3$ concentration in the MLT is five orders of magnitude lower than that of $O_2$.

In principle, reaction (9) is endothermic and will proceed rapidly in the hot meteor boundary region. When M = $Fe^+$ or $Mg^+$ in the equations above, observational evidence indicates that under hyperthermal conditions, subsequent reactions proceed at the collisional rate (Ferguson & Fehsenfeld 1968). Furthermore, the rate of the reaction depends on thermodynamic and mixing considerations in the boundary region between the shock modified ambient atmosphere and the metal ions produced in the ambipolarly diffusing meteor trail (Dressler 2001; Jenniskens et al. 2004; Berezhnoy & Borovička 2010).

The formation of $MO^+$ will take place between 3,000 K and 1,500 K (for additional discussion on metal oxide formation in meteor trains at temperatures between 1,500 and 4,000 K see

Berezhnoy & Borovička 2010), which indicate a reasonable range of values of temperature in the meteor train boundary for the first 0.1s (during the initial stage of the ambipolar expansion) (Jenniskens et al. 2004). The process of production of a metal oxide ion by the hyperthermal reaction between $M^+$ and $O_2$ yields highest quantities at about 2,500 K and it is not appreciable below 800 K, which is the value below which oxygen is inert (Zel'dovich & Raizer 2002). In the lower temperature range, this process will cease to be relevant as the source of the metal ion oxides. Observational data of meteor wake temperatures and train thermalization (e.g. Jenniskens et al. 2004) supports our assertion that the metal oxide ion formation takes place following the adiabatic overdense meteor train 'instantaneous' expansion.

However, the thermochemistry of these reactions is poorly established (Dalleska & Armentrout 1994), especially under the MLT conditions. Armentrout, Halle & Beauchamp (1982) investigated the reactions of $Cr^+$, $Mn^+$, $Fe^+$, $Co^+$, and $Ni^+$ with $O_2$, which yield metal oxide ions and reported on the reaction cross sections as a function of ion kinetic energy. A number of studies have been conducted in the past investigating the thermal reactions between metals, such as Fe, Mg, Al, and $O_2$ (e.g. Fontijn & Kurzius 1972; Fontijn et al. 1972). However, only a relatively small number of studies considering the metal ion reactions with oxygen have been performed (e.g. Armentrout et al. 1982).

Subsequently, the removal of electrons by the thermally formed meteor metal oxides (produced in the postadiabatically diffusing train boundary) is an exothermic process that is both fast and temperature independent for meteor metal oxide ions (Plane et al. 2015):

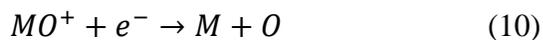

$$MO^+ + e^- \rightarrow M + O \qquad (10)$$

The time scale of electron removal during this reaction depends on the number density of the newly formed $MO^+$. Following the consumption of a critical number of meteor metal oxides, the reaction (10) will no longer be appreciable and then ambipolar diffusion takes over again (Figure 1) as a dominant mechanism of electron removal from the meteor train. Depending on the altitude, the process of electron removal will be complete in approximately 0.1 - 0.3 seconds. The above discussion indeed demonstrates that ozone (albeit indirectly) may play a role in a brief electron removal from the postadiabatically expanding meteor train boundaries. At the moment, in the absence of high resolution numerical code that accounts for the shock induced chemical reactions of both major and minor MLT species in the rarefied atmosphere, we cannot estimate with certainty the ratios of shock dissociated ozone and ambient shock surviving $O_2$ that participates in the initial thermally driven oxidation and subsequent post-hyperthermal dissociative recombination that removes electron from meteor train boundary (reaction 10) (See the comments in Section 4.3). We can say that in the volume between $r_0$ and $R_0$, however, most of $O_2$ that originates from shock dissociation of $O_3$ will be consumed by meteor metal ions. We can, however, estimate that the ratio of $O_2$ that comes from ozone shock dissociation to the ambient $O_2$ that participates in thermally driven oxidation of meteor metallic ion is in the range of $10^{-2} - 10^{-3}$, depending on the axial distance from the meteor train (within $R_0$). This is still a considerable contribution from ozone, given that the ratio of $O_3$ to $O_2$ in MLT is about $10^{-5}$.

In summary, it has been demonstrated that an initial hyperthermal chemical reaction (where the rate is governed by the temperature), which is subsequently followed by a dissociative recombination (that will primarily depend on the concentration and availability of metal oxide ions) may be instrumental in removing electrons from the postadiabatically expanding high temperature meteor trail. Moreover, both of these processes are competing against ambipolar diffusion. The best way to describe such dynamic system is with the second Damköhler number ($Da_{II}$) (Sarma 2000). Consequently, the electron removal in the boundary region of the ambipolarly expanding high temperature meteor train will strongly depend on the second Damköhler number, which needs to be always considered when there are competing regimes of diffusion and chemical removal of electrons in the meteor train boundary region (Jakobsen 2008; Jarosinski & Veyssiere 2009).

We note that $Da_{II}$ is a function of ambipolar diffusion, temperature and species number density. For $Da_{II} > 1$, the chemical removal of electrons dominates, while for $Da_{II} < 1$, the ambipolar diffusion is a primary mechanism of electron removal.

Of course, the process of the electron removal discussed above may not be applicable farther away ($\geq R_0$) from the boundary of the initially formed meteor train where the effects of the cylindrical shock waves attenuate rapidly. Specifically, we have shown in this section that the overdense meteor cylindrical shock wave-dissociated ozone products (ground state $O_2$ (X) and to a lesser degree $O_2(a^1\Delta_g)$) play an important role through hyperthermal reactions with meteoric metal ions and subsequent dissociative recombination in electron removal from the boundary of the overdense meteor train in the early stages of the postadiabatic trail expansion.

Interestingly, the species densities and evolution in the early stage of meteor trail boundary evolution, modeled by Zinn et al. (2004) and Zinn & Drummond (2005), support our findings.

Finally, the importance of hyperthermal chemistry enabled by the relatively slow thermalization of the high temperature meteor train (Jenniskens et al. 2004) and likely modification of the nearfield region of the ambient atmosphere, even by a relatively weak shock wave, can be further extended not only to transitional, but also to strong underdense meteors (e.g. see Lee et al. 2013; and Hocking et al. 2016).

**4.3 The Computational Model**

For illustrative purposes, we have modeled a simple hypersonic meteor flow without ablation in the MLT region. The broad aim was to emphasize the difference non ablation makes relative to the ablating meteoroids, and make a qualitative comparison to those numerical models which do include ablation (e.g. Boyd 2000; Zinn et al. 2004). We applied a simplified model, incorporated into the computational fluid dynamics (CFD) software package ANSYS Fluent (http://www.ansys.com), to investigate the distribution and magnitude of the pressure and temperature fields behind the initial bow shock wave envelope, and to determine what fraction of $O_2$ (if any) survives the initial meteor bow shock wave conditions. The computational model, along with the governing equations and rate parameters (based on Park (1989)), are described in Niculescu et al. (2016). The code is optimized for simulating the formation and evolution of the bow shock wave in the continuum flow regime associated with the hypersonic flows and the model can resolve the chemical reactions of the major species in and behind the shock wave (Niculescu et al. 2016). Shock waves, as well as the chemical reactions, including the dissociation of $N_2$ and $O_2$ are included in the model. However, the code does not currently include modeling certain minor species, such as $O_3$. At the moment, this simple model is not optimized to resolve the effects of ablation, ionization and radiation. However, efforts are being made at the moment to incorporate those effects in future numerical simulations.

The computation was performed using $O_2$ and $N_2$ as the only major species, at an altitude of 80 km. Relative to the ambient air, the initial mass fractions are 0.233 ($O_2$) and 0.767 ($N_2$), and the initial molar fractions are 0.21 ($O_2$) and 0.79 ($N_2$). A spherically shaped meteoroid is assumed to be moving at 35 km/s ($M_{80km}$ = 124.6). The meteoroid diameters ($m_d$) considered in our simulations are $m_d$ = 2.5 cm and $m_d$ = 10 cm. The ratio of hydrodynamic to chemical time scale of $O_2$ formation and destruction for these two cases is 0.001 and 0.01, respectively. Thus, we used the non-equilibrium approach.

The computational results representing the mass fraction of $O_2$, pressure, and temperature fields are shown in Figures 3 – 5. The radial temperature distribution is plotted in Figure 6.

We have used the simulation results obtained here to infer the amount of ambient $O_2$ that will survive the passage of the cylindrical shock wave (Figure 3). Although the effect of dissociation on temperature in the flow field is included in the computational model, the present version of the code does not allow for a precise estimate of the amount of energy dependent dissociation behind the cylindrical shock wave. Subsequent improvements in the code are needed to cover that aspect. Nevertheless, it is possible to infer with reasonable certainty that under the meteor cylindrical shock wave conditions (discussed in the main text, and Sections S1 and S2), not only will a significant amount of $O_2$ survive, but some $O_2$ that comes from $O_3$ dissociation will also persist.

Since the effect of ablation is not included in the model, the magnitude of the pressure (Figure 4) in the flow field is correspondingly smaller (e.g. Bronshten 1983). The recompression (cylindrical) shock wave can be seen forming and its magnitude and effects are negligible in comparison with the initial bow shock wave or the cylindrical shock wave in the case of ablation. In the absence of ablation, the size of the flow field behind the initial shock region in front of the meteor is up to two orders of magnitude smaller than for the case of ablation (Boyd 2000; Jenniskens et al. 2000; Zinn et al. 2004). While there is a significantly smaller pressure increase behind the initial shock from the non-ablating spherical object, and overall reduced size of flow fields, the temperature magnitude (Figure 5) remains reasonably similar to the case where ablation is considered (see Zinn et al. 2004 and Boyd 2000). The magnitude of the temperature (Figure 6) is relatively similar to a scenario where strong ablation is present (Boyd 2000; Jenniskens et al. 2000; Zinn et al. 2004). However, the absence of the ablation will significantly reduce the magnitude of the radius of the volume around the meteor axis with an increased shock velocity dependent temperature. Although the effect of radiation is not included, our results (in spatially scaled down version) are consistent with those presented in Zinn et al (2004).

While the numerical model of the temperature and flow fields, along with the appearance of the bow shock wave, is in line with observations and other theoretical results (e.g. Viviani & Pezzella 2015), we have demonstrated that in the absence of ablation, the flow regime remains unaffected, as predicted theoretically for the bodies of the specified sizes (Boyd 2000; Jenniskens et al. 2000; Zinn et al. 2004). Moreover, the recompression shock wave is substantially weaker than the radially expanding initial bow shock envelope, especially in comparison to the models that include ablation (Boyd 2000; Jenniskens et al. 2000) and observational results (Jenniskens et al. 2004). However, in the near field the radially expanding bow shock will still have modifying effects on the narrow region around the hypersonic body.

## 5. Summary and Conclusions

In this work we have examined and presented the link between overdense meteor generated shock waves and the short lasting hyperthermal chemistry regime during the initial evolution of the meteor train. From the theoretical considerations, our results and conclusions can be summarized as follows:

i. Ablationally amplified cylindrical shock waves (approximated as blast waves from an explosive line source) produced by overdense meteors are strong enough to substantially modify the ambient atmosphere in the region near the initial point of maximum energy deposition per unit path length. The average overdense meteor cylindrical shock wave (which directly depends on pressure) heats the ambient atmosphere to about 6,000 K in the near-field region ($<R_0$). This theoretical calculation is based on determinations of both (i) the meteor velocity and delivered energy and (ii) the pressure ratio between the ablated and entrained vapour and plasma in front of the propagating meteoroid relative to the ambient atmosphere pressure. A temperature in the range of 6,000 K is sufficient to dissociate both ozone and oxygen.

ii. Specifically, we have demonstrated that in the range of initial temperatures in the region behind the strong overdense meteor cylindrical shock wave and in the meteor trail near-field (within $R_0$), large quantities of $O_2$ will be dissociated. On the other hand, large quantities of both ground level and to a lesser extent, excited $O_2$, which originate from ozone shock dissociation, survive. This is primarily due to the finite energy budget and short time scales in the meteor region of MLT. However, both shock surviving ambient oxygen and that originating from $O_3$ dissociation, hyperthermally react with the meteoric metal ions in the boundary region of the high temperature postadiabatically expanding overdense meteor trail. The time scales for high temperature oxidation of meteoric metal ions depend strongly on the temperatures at the meteor boundary and altitude, and are typically on the order of approximately $10^{-3}$ s at 80 km.

iii. Furthermore, for the case of overdense meteor trains, we have demonstrated that the subsequently formed meteoric metal oxide ions are predominantly responsible for the initial intense and short lasting electron removal from the boundary of the expanding meteor train, through a process of fast temperature independent dissociative recombination. This height dependent process is generally completed within 0.1- 0.3 s, which agrees well with the results indicating significantly slower cooling of meteor wakes (Jenniskens et al. 2004). The rate of this process is also strongly dependent on the second Damköhler number.

Finally, the findings presented in this paper are significant as they illuminate the combined role of previously neglected effects of meteor generated shock waves and hyperthermal chemistry in the role of radar-observed early diffusion of electrons in the meteor train boundary, which are consumed in the posthyperthermal dissociative recombination. Evidently, there is a need for further validation using not only more capable numerical models, but also additional observational and experimental studies.

**Acknowledgements:**

EAS gratefully acknowledges the NSERC Postdoctoral Fellowship program for partly funding this project. WKH acknowledges a Discovery Grant from the Natural Sciences and Engineering Research Council of Canada. MG acknowledges support from the ERC Advanced Grant No. 320773, the Russian Foundation for Basic Research, project nos. 16-05-00004 and 16-07-01072, and the Russian Science Foundation, project No. 14-22-00197. RES thanks NSERC CREATE Training Program for Integrating Atmospheric Chemistry and Physics from Earth to Space (IACPES), and a Northern Scientific Training Program grant. Research at the Ural Federal University is supported by the Act 211 of the Government of the Russian Federation, agreement No 02.A03.21.0006.

## Figures and Figure Captions:

**Figure 1:** Schematic depiction of an overdense meteor's early evolution, in which three distinct stages can be recognized. In the first stage, the ablating meteoroid with the shock front in front sweeps the cylindrical volume of ambient atmosphere (depicted by the small gray circle), ionizing and dissociating atmospheric gasses. This stage also coincides with the cylindrical shock wave expanding radially outward, perpendicular to the meteor axis of propagation, with enough energy deposited within $R_0$ to dissociate $O_2$ and $O_3$ in the ambient atmosphere, but not enough for $N_2$ dissociation (see the main text for discussion). In stage two, the adiabatically formed meteor train (which can be approximated as quasineutral plasma with the Gaussian radial electron distribution), begins to expand under ambipolar diffusion and thermalizes. This stage coincides with formation of metal ion oxides which takes place and is appreciable between approximately (3,000 – 1,500 K) at the boundary region of the diffusing trail. In this reaction, an ablated meteoric metal ion will react in a thermally driven reaction with the shock-dissociated product of ozone ($O_2$ in ground and excited states). In the stage three, in the almost thermalized train, the newly formed metal ion oxide will consume electrons rapidly by temperature independent dissociative recombination (see the main text for discussion).

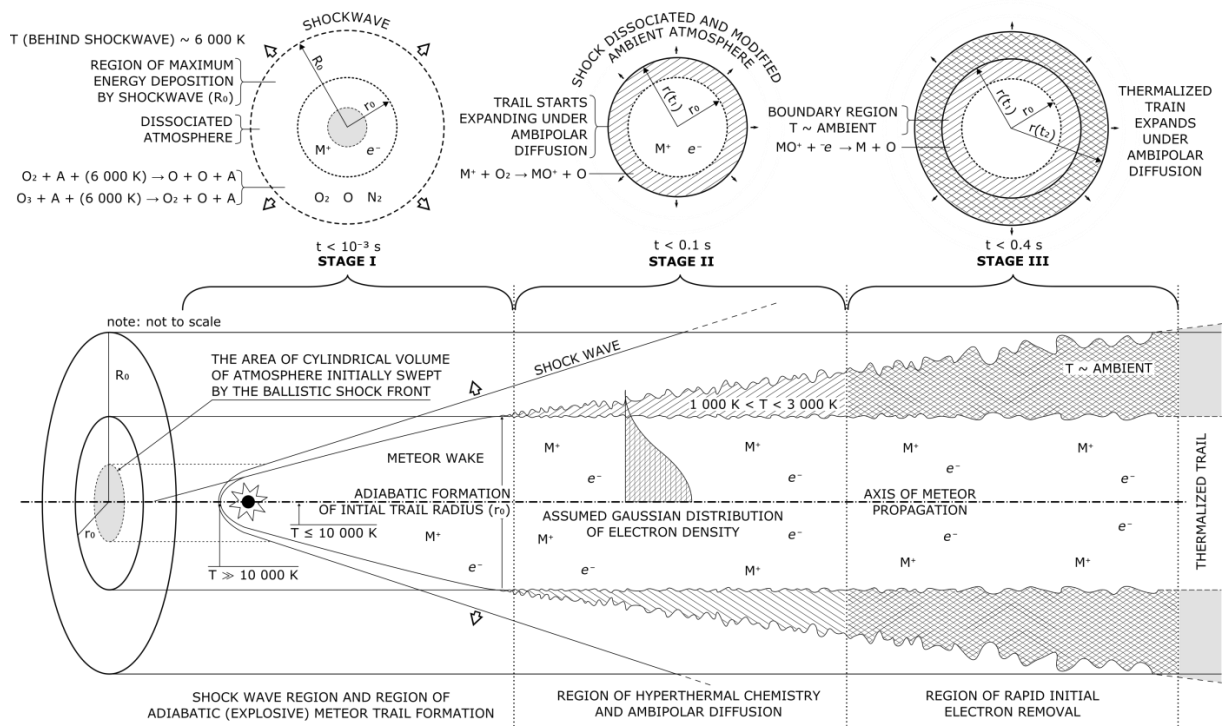

**Figure 2:** (a) Plotted are the initial radius ($r_0$) of a typical bright overdense meteor (from Baggaley & Fisher 1980) and the radius of the meteor ($r_m$) trail after $t = 0.3$ s. These are compared to $R_0$ as a function of constant energy deposition (see eq (1)) of 100 J/m and 1000 J/m for altitudes from 80 km to 100 km. For $r_m$ at 0.3 s, we applied the geometrically averaged hot plasma and ambient atmosphere ambipolar diffusion coefficients as per Hocking et al. (2016), appendix C. (b) The initial radius $r_0$, plotted along $r_m$ at $t = 0.3$ s. Shown here is the comparison between $r_m$ as calculated in panel (a), and $r_m$ as calculated using the Massey's formula for the theoretical diffusion coefficient (Jones & Jones 1990).

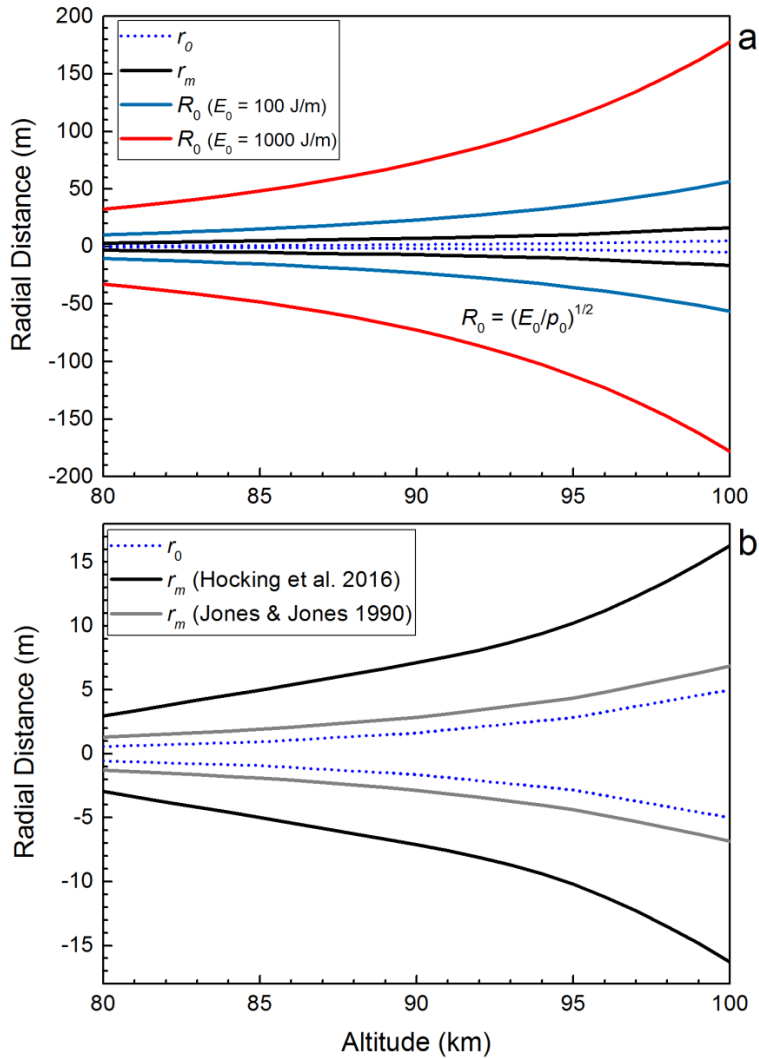

**Figure 3:** The mass fraction of $O_2$ as a function of radial distance from the propagation axis of the (a) 2.5 cm and (b) 10 cm meteoroid. The top boundary ("white space" in the plot) represents a numerical boundary condition without any physical significance (it is set up to be far enough away from the body (meteoroid), such that the influence of the body (meteoroid) no longer has any effect. Note that (a) and (b) have different axes scaling. The colour scheme is represented in log scale.

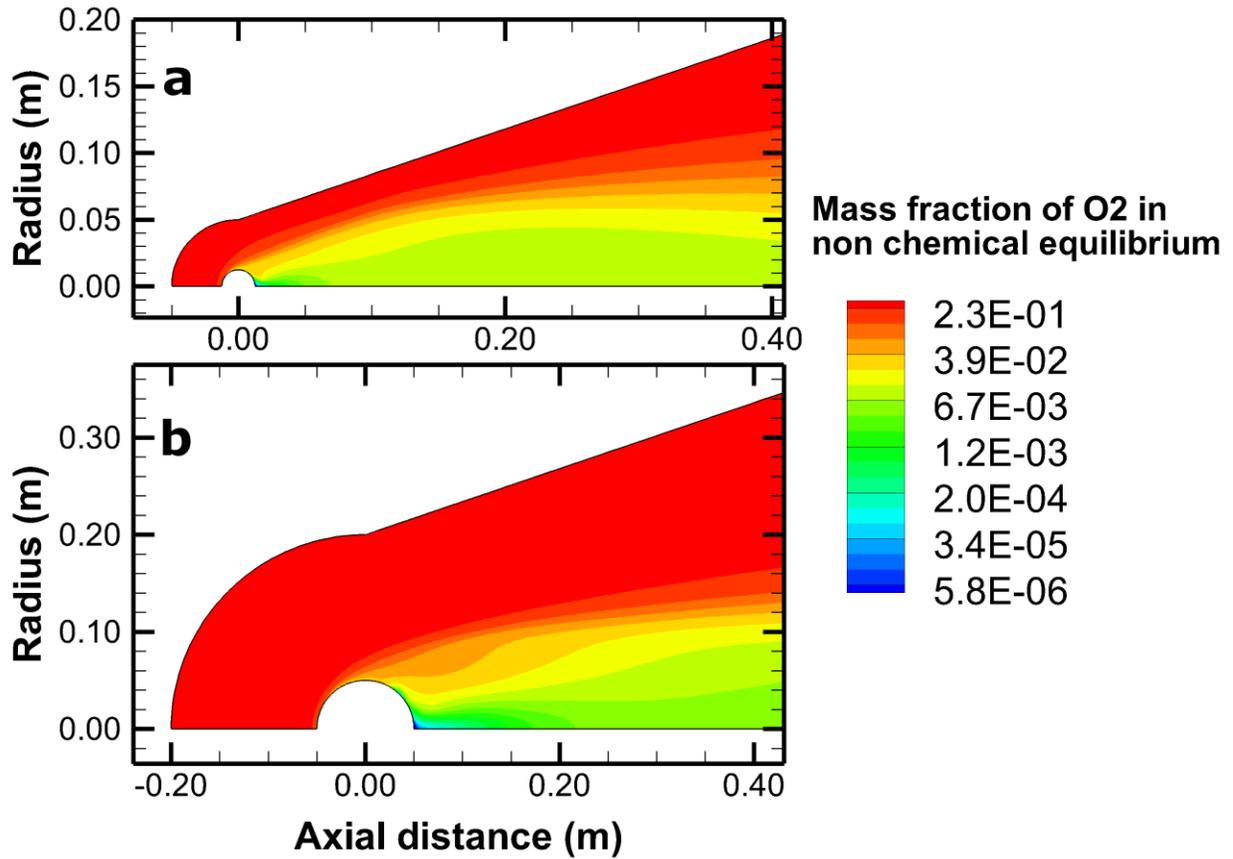

**Figure 4:** The pressure distribution around the propagation axis of a (a) 2.5 cm and (b) 10 cm meteoroid. The top boundary ("white space" in the plot) represents a numerical boundary condition without any physical significance (it is set up to be far enough away from the body (meteoroid), such that the influence of the body (meteoroid) no longer has any effect). Note that (a) and (b) have different axes scaling.

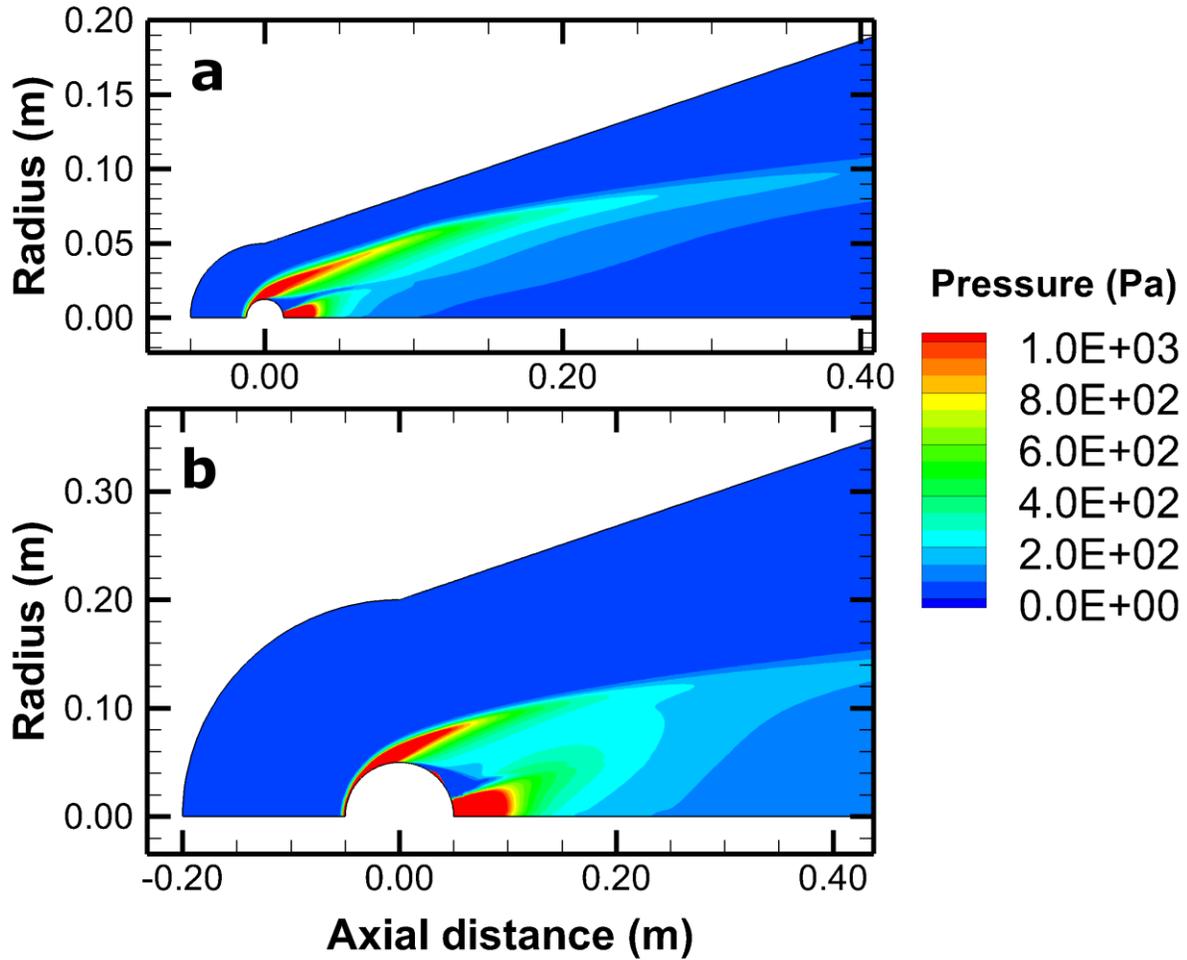

**Figure 5:** The temperature distribution around the (a) 2.5 cm and (b) 10 cm meteoroid. The top boundary (blank region in the plot) represents a numerical boundary condition without any physical significance (it is set up to be far enough away from the body (meteoroid), such that the influence of the body (meteoroid) no longer has any effect). Note that (a) and (b) have different axes scaling.

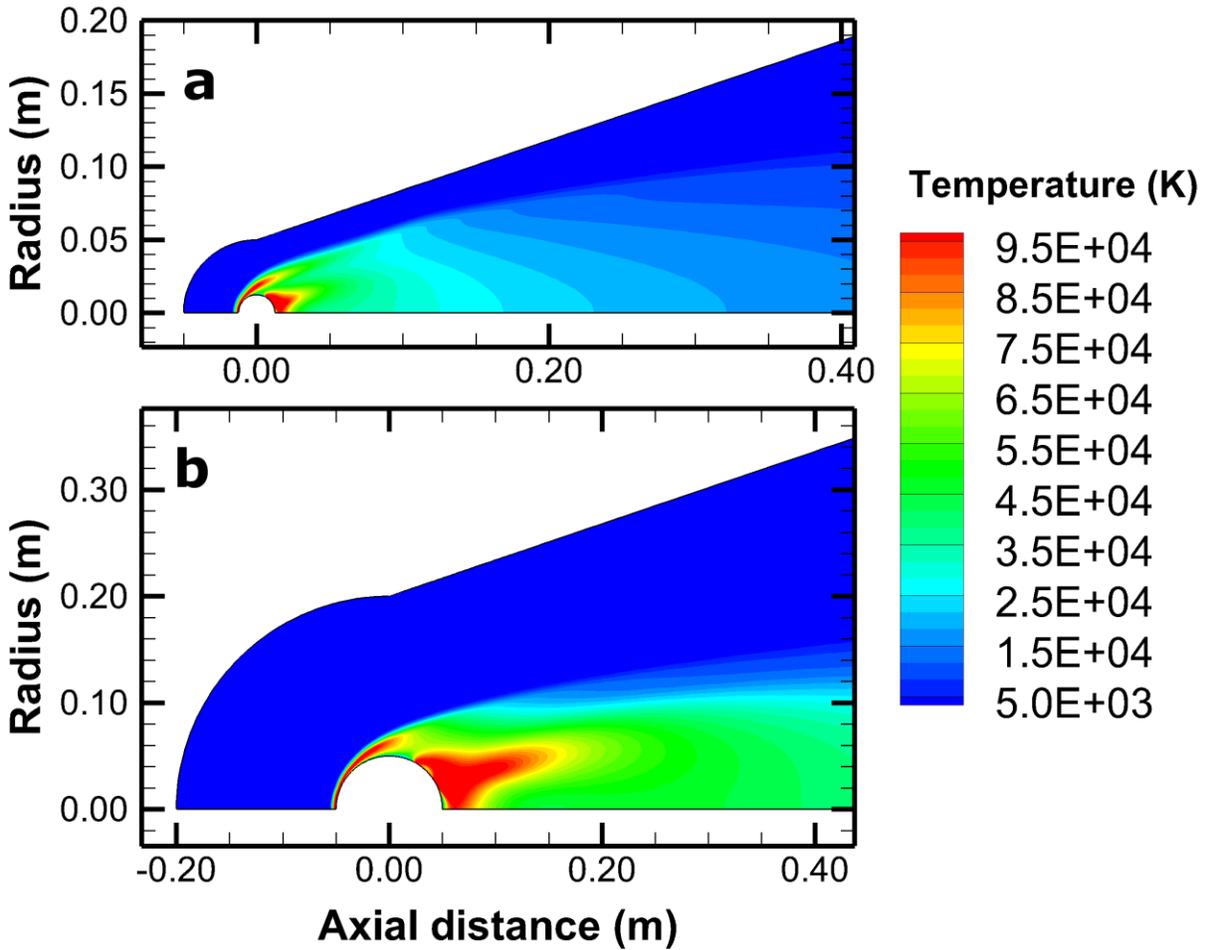

**Figure 6:** The radial temperature distribution as a function of distance from the meteor (the vertical axis is in log scale).

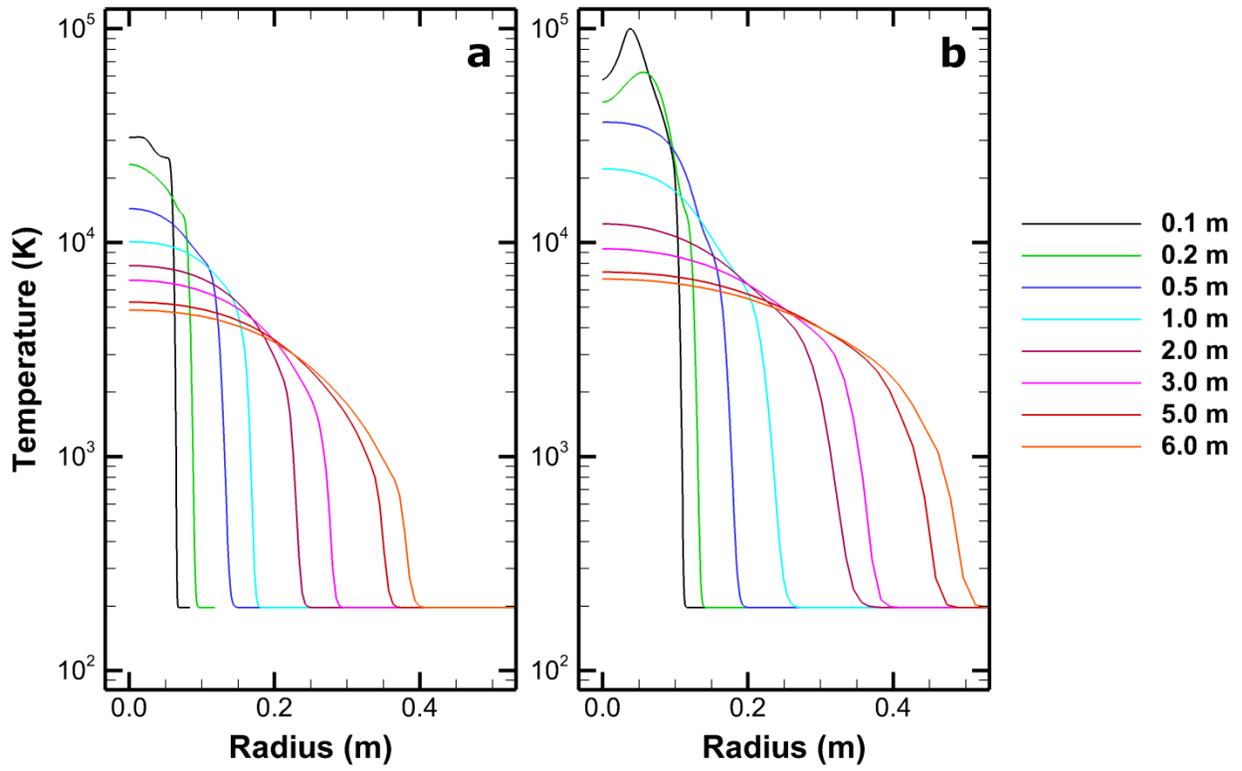

# Supplementary Material for

# On Shock Waves and the Role of Hyperthermal Chemistry in the Early Diffusion of Overdense Meteor Trains

Elizabeth A. Silber[*,1], Wayne K. Hocking[2], Mihai L. Niculescu[3], Maria Gritsevich[4,5,6], Reynold E. Silber[7]

[1] Department of Earth, Environmental and Planetary Science, Brown University, Providence, RI, 02912, USA

[2] Department of Physics and Astronomy, University of Western Ontario, London, Ontario, N6A 3K7 Canada

[3] INCAS - National Institute for Aerospace Research "Elie Carafoli", Flow Physics Department, Numerical Simulation Unit, Bucharest 061126, Romania

[4] Department of Physics, University of Helsinki, Gustaf Hällströminkatu 2a, P.O. Box 64, FI-00014 Helsinki, Finland

[5] Department of Computational Physics, Dorodnicyn Computing Centre, Federal Research Center "Computer Science and Control" of the Russian Academy of Sciences, Vavilova St. 40, 119333 Moscow, Russia

[6] Institute of Physics and Technology, Ural Federal University, 620002 Ekaterinburg, Russia

[7] Department of Earth Sciences, University of Western Ontario, London, Ontario, N6A 3B7, Canada

**Section S.1.**

**S.1 Knudsen Number, Flow Regimes and the Formation of Shock Waves**

With regard to meteors, we consider hypervelocity flow as referring to the flow of atmospheric gas over the meteoroid when the Mach number, $M_\infty$ (defined as the ratio of the flow velocity to the local speed of sound), ranges between 35 and 270 (e.g. Boyd 2000; Silber et al. 2015). The main difficulty in treatment and analysis of meteoric hypersonic flows and the resulting bow and cylindrical shock waves is the highly nonlinear nature of the problem. In hypersonic gas dynamics (see, for example, Zel'dovich & Raizer 2002; Anderson 2006) any hypersonic blunt body (a meteoroid can be approximated as such, with some refinements, to be discussed in the next paragraphs), will form a strong bow shock wave front, and a flow field between the shock wave and the body - generally referred to as the shock layer - will develop (Anderson 2006). For near-spherical bodies the thickness of the shock layer can be approximated using the compression coefficient $\rho/\rho_s$ as $l=0.8R\rho/\rho_s$ (Stulov 1969; Gritsevich et al. 2011). Here, $\rho$ and $\rho_s$ are densities of the impinging flow and gas behind the shock wave, respectively, $l$ is the layer thickness and $R$ is the radius of the sphere. For hypersonic entry vehicles (e.g. the case of ballistic and re-entry vehicles), the existence of this phenomena has been known since the mid-last century, where the shock layer is relatively thin and depends on the type of the flow regime. The similarity parameter that describes the type of flow is the Knudsen number ($Kn$), defined as the ratio of the local atmospheric mean free path and characteristic dimensions of the body (in our case, the meteoroid diameter). Larger vales of $Kn$ refer to smaller bodies. Four basic types of flow regime can be defined. The free molecular flow is defined for $Kn > 10.0$, and a transitional flow regime exists when $0.1 < Kn \leq 10.0$. A Knudsen number satisfying $0.01 \leq Kn \leq 0.1$ signifies the slip flow regime, and values of $Kn$ below 0.01 indicate continuous flow (Josyula & Burt 2011). The first three flow regimes, respectively, are important in rarefied gas dynamics.

However, for an ablating meteor, the matter is complicated by the existence of a vapour cloud in front of the meteoroid, the size of which is proportional to the cube of the meteoroid velocity (Öpik 1958; Bronshten 1983; Popova et al. 2000; Boyd 2000; Campbell-Brown & Koschny 2004). The vapour cloud is formed as the incoming meteor collides with the atmospheric molecules and atoms, producing reflected atmospheric atoms, along with large numbers of scattered evaporated and ablated meteoric constituents. The reflected and evaporated constituents of this cloud may have velocities up to $1.5v_{meteor}$ (Rajchl 1969). In such cases, the shielding effect of the cloud becomes important for most meteors (Popova et al. 2000; 2001). The vapour cloud formation in turn initiates aerodynamic and thermal shielding, where the role of convective and radiative heat transfer during the meteoroid ablation becomes far more significant, especially following the formation of the shock wave (Öpik 1958; Bronshten 1983).

For example, Boyd (2000) calculated that each impinging air molecule will release about 500 meteoric particles, and the rate of ejection is a function of not only meteoroid velocity but also the composition. This, in addition to producing very high-temperature chemically reactive flows, will affect the dimensions of the aerodynamic shielding, shock layer and subsequently the boundary layer (Popova et al. 2000; 2001; Boyd 2000; Jenniskens et al. 2000; Anderson 2006). Rajchl (1969) also considered the vapour cloud shielding in front of the meteor and determined that it may affect the ionization efficiency coefficient and consequently the number density of free electrons in the meteor trail. The appearance of the vapour cloud occurs in the lower region of the free molecular flow regime during the transition into the slip-flow regime (Bronshten 1965; Popova et al. 2000). Thus, in the case of the meteoroid transit thorough the atmosphere, the flow conditions are also determined by the mean free path of the reflected molecules relative to the impinging molecules in a frame moving with the particle (Bronshten 1983).

The mean free path of reflected molecules and atoms in the vapour cloud is also a function of Mach number, local mean free path and meteor-atmosphere temperature ratio, and can be expressed in those terms (see Bronshten 1983).

Most meteoroids ablate in the rarefied region of the atmosphere between 70 km and 120 km. At these altitudes, the mean free path of the ambient atmosphere varies by up to ±2 orders of magnitude relative to the meteoroid dimensions. The inclusion of the specific flow regime in the analysis of the meteoroid interaction with the atmosphere is of critical importance as it will affect the analytical and computational treatment of ablation and mass loss, ionization (Campbell-Brown & Koschny 2004), shock waves, and other relevant physical parameters (see Bronshten 1983; Popova et al. 2000). Popova et al. (2000) simulated density increase in the 'vapour layer' in front of the meteoroid. They obtained values more than two orders of magnitude higher than the density of the local region of the atmosphere. It should be noted that the size of the modeled vapour cloud, obtained by existing computational methods, is relatively small in comparison with high resolution observations (Jenniskens & Stenbaek-Nielsen 2004; Stenbaek-Nielsen & Jenniskens 2004). Additionally, the assumed ideal gas specific heat ratio used in any simulation will affect the size of the vapour cloud, since it determines the thickness of the boundary layer. The formation of the vapour cloud will indeed alter the flow regime considerations (Boyd, 2000; Popova et al. 2001), shifting the free-molecular flow to higher altitudes and thereby affecting both the Knudsen number and Reynolds number (the ratio of inertial forces to viscous forces in a fluid flow) (e.g. Anderson 2006). This will subsequently affect the timing and formation of the meteor shock wave front (Jenniskens et al. 2000). For $Re \leq 1$, viscous forces must be considered, whereas for $Re \geq 1$, the viscosity is can be neglected. In principle, overdense meteor shock waves will be relevant below 95 km, noting again the dependence on the meteoroid size, velocity, composition and ablation.

An important observation by Popova et al. (2000) indicates that below an altitude of 114 km, the pressure of the vapour cloud in front of the meteoroid will be always greater than the ambient air pressure for all bodies with sizes 0.1 mm – 100 mm. The implication is that for most overdense meteors, the continuum flow applies below approximately 90-95 km, while the transitional and strong underdense meteors will still reside in the transitional regime below 80-85 km in altitude (Popova et al. 2000).

The transition from the vapour cloud to the ballistic shock front occurs with the meteoroid's descent to the lower altitudes (lower transitional flow regime) when the vapour pressure, density and temperatures are much higher than that of the ambient air (essentially producing a discontinuity at the edge of the cloud), at which point the vapour can be treated like a hydrodynamic flow into a vacuum (Popova et al. 2000; 2001).

The flow within the shock layer can be appropriately described by the Navier-Stokes equations for compressible flow (Hayes & Probstein 1959). The discontinuity satisfies the Rankine-Hugoniot relations (e.g. Sachdev 2004; Vinnikov et al. 2016) (which relate the upstream and downstream values of density, bulk velocity, and temperature in an ideal compressible fluid), and leads to the formation of a compressed shock layer and viscous boundary region close to the body (Probstein 1961; Bronsthten 1983). Probstein (1961) indicated that the shock wave is *efficient* when the depth of the shock front ahead of the body corresponds to the radius of the body and when the mean free path behind the shock is one third of the meteoroid radius.

It has been demonstrated that the meteor shock waves appear much earlier than predicted by classical gas dynamics theory. This occurs before the onset of the continuum flow (e.g. Bronshten 1983; Probstein 1961; Popova et al. 2001) and for most meteoroids will take place in the lower region of the transitional flow regime. In principle, all overdense meteors will exhibit strong shock wave at altitudes where they ablate; the governing parameters are presented by Probstein (1961).

**Section S.2**

**S.2 Meteor Cylindrical Shock Waves**

In principle, any hypersonic body (e.g. re-entry vehicle) propagating through a gas, with velocity higher than the local speed of sound in the gas, will produce a shock wave. The shock wave generated by a much faster meteoroid consequently propagates at much higher hypersonic velocities and can be considered a shock from an explosive line source (Lin 1954; Tsikulin 1970). The hypersonic nature of the initial meteor shock front (which transitions to a bow shock wave behind the meteoroid, and envelopes the ablating vapour volume, extending in a parabolic or sometimes approximated as the cylindrical shape) will yield a high degree of ionization and dissociation for both ablated meteoric atoms and colliding atmospheric molecules. Note that while the meteor bow (and cylindrical) shock wave propagates hypersonically in the radial direction into the surrounding atmosphere, normal to the meteoroid propagation axis, the initial volume of high temperature vapour and plasma(Zinn, O'Dean & ReVelle 2004) enclosed by the bow shock behind the transiting meteor expands rapidly and adiabatically, attaining a finite volume with an initial radius $r_0$ (Jones 1995), after which the expansion is governed by the ambipolar diffusion regime (after plasma thermalization).

Let us consider a typical overdense meteor (considered as a blunt body with a spherical shape) interacting with the MLT atmosphere in the transitional flow regime.

In such a simplified scenario, when the viscous and thermal effects are neglected, the governing equations for the flow in cylindrical coordinates are given by Plooster (1970) and Tsikulin (1970). The comprehensive cylindrical blast wave theory is also discussed by Sakurai (1964) . The velocity of the cylindrical shock front will rapidly attenuate and transition to the acoustic regime within $10R_0$. Here, $R_0$ is the characteristic radius of the maximum energy deposition by an ablating hypersonic meteoroid, per unit path length (Tsikulin 1970), and differs from $r_0$.

Following the transition from the condensed vapour cloud (Popova et al. 2000) to the shock layer ahead of the meteoroid, which forms due to a sudden decrease of the flow velocity near the surface, it is possible to define distinct features of the meteor flow field and the shock wave. Figure S1 shows a schematic representation (albeit not to scale) of a hypersonic meteoroid interacting with the atmosphere, emphasizing key regions such as the flow field and shock wave(s). The ablation amplified bow (cylindrical) shock wave front (1) that envelops the meteoroid (7) is the extension of the shock wave front (2) ahead of the meteoroid. The ballistic shock front formed in front of the meteoroid consists of the detached shock front, stagnation region, and viscous boundary layer described below.

The shock front (2) is fundamentally a thin region with very strong gradients of pressure, kinetic temperature and density which results from continued adiabatic compression of the vapour cloud in front of the ablating meteoroid. The thickness of the shock wave front in the rarefied atmosphere is on the order of one mean-free-path (Taniguchi et al. 2014). However that thickness is a function not only of the ratio of specific heats, but also of the dissociation rate of atmospheric molecules (Zel'dovich & Raizer 2002) and in the case of a meteoroid, of its ablation rate.

It is important to note that, in the analysis of the meteor shock wave (including the ballistic shock front), the assumption of a perfect gas is not valid (Steiner & Gretler 1994). The equation of state must take into account dissociation, electronic excitation and ionization.

Within the front shock layer, characterized by an almost instantaneous increase in temperature, density and entropy (Anderson 2006), we can distinguish several distinct features, as described in the following paragraphs.

Behind the bow shock front appears the sonic region (3), where the flow velocity is decelerated to subsonic values. The viscous boundary layer (4) envelops the meteoroid and is characterized by non-equilibrium processes (Zel'dovich & Raizer 2002; Anderson 2006), where the thickness will depend on the mean free path, chemical reactions, and the ratio of the specific heats. The boundary layer also hosts the stagnation point (5) (or the stagnation region, depending on the size of the object) which is the region with the highest translational temperatures. This region is where energy is converted to radiation (Bronshten 1983) and where most of the meteor UV radiation is produced.

The boundary layer extends behind the meteoroid and evolves into a 'free' shear layer (9), characterized by differences in velocities between the outer and inner flow, where the ablated meteor plasma and vapour, mixed with the entrained dissociated and ionized atmospheric constituents, is streamed toward the recompression region or the neck (8). The neck (8) is a region of high-pressure compressed flow and very high temperature (generally source of the recompression shock or, cylindrical shock waves in the case of strongly ablating meteoroids). The recirculation zone (6) (in older literature referred to as the 'dead water region') is driven by high pressure at the wake neck, and flows back toward the aft end of the body and outward toward the shear layer separation point (Gnoffo 1999).

The propagation and evolution of the ablated and evaporated material immediately around and in the meteor wake can be described by the Navier-Stokes equations for continuity, momentum and energy, with the first two being purely mechanical and not affected by chemical processes (Anderson 2006). However, in the remaining energy equation a consideration must be given to viscous, thermal and chemistry effects in the hypersonic gas flow (Anderson 2006). In the neck-region, the ablated vapour and plasma are compressed and recompressed behind the meteoroid and are described by two velocity components; one parallel to the axis of the meteoroid motion and other perpendicular to it. This is where the secondary recompression shock wave (10) is formed (also referred to as the ablational shock wave) which is significantly stronger (for the case of strongly ablating meteoroids) than the outer bow or primary cylindrical shock wave (Dobrovol'skii 1952). This is due to extremely high temperatures ($T > 10,000$ K) and pressures (relative to the ambient pressure) of the flow field. However, while it is possible, for pedantic reasons, to separate the two cylindrical shock waves, ultimately both rapidly converge into one cylindrical shock wave and can subsequently no longer be distinguished (Hayes & Probstein 1959). Vapour and plasma in the meteor wake (11), with initially strong turbulence (Lees & Hromas 1961) expand rapidly and adiabatically to form the meteor train initial radius within $10^{-3}$ s, observable by radar.

The hypersonic and chemically reactive flow in the transitional flow regime is generally considered a non-equilibrium flow because the distribution of energies among internal modes and the distribution of species in any small parcel of gas in the flow field are functions of the collisional history of those molecules (Gnoffo 1999). Sarma (2000) presented a comprehensive discussion and illustrated the complexity of the physico-chemical processes in the hypersonic flow.

The temperature behind the shock wave in real gases (such as is the case in the meteor region) will be always less than that predicted by the theory for calorically perfect ideal non-reacting gases.

In the region behind the shock front (in this case we consider the ballistic shock front ahead of the meteoroid) the translational temperature rises rapidly and then falls off quickly in the subsequent flow field as the energy is transferred to rotational temperature. The equilibrium between the translational and rotational modes is established rapidly. The vibrational energy on the other hand rises more slowly; the equilibrium between all three modes is reached after the coupled loss of translational and rotational energies (Hurle 1967).

In principle, the physico-chemical effects behind the strong detached (ballistic) shock front are well defined in the meteor literature (e.g. Bronshten 1965; Menees & Park 1976; Park & Menees 1978; Berezhnoy & Borovička 2010). The region bound by the initial bow shock envelope will be completely dissociated and ionized, and will contain the highest concentration of meteoric metal ions. The reason for this is that the main meteor metals (Fe, Mg, Al, Ca) have correspondingly much lower ionization potentials compared to the atmospheric molecules and atoms (e.g. $O_2$, $N_2$, O, N) (Dressler 2001). The hyperthermal chemistry, such as the formation of nitric oxide, will take place as long as temperatures are adequately high (Zel'dovich & Raizer 2002) and will cease when the temperature falls below some critical value (Park & Menees 1978).

In this rarefied region of the atmosphere, the strongest effects of the meteor cylindrical shock wave(s) are close to the source; they occur within $R_0$ of the initial maximum energy release region, and the shock wave is sufficiently attenuated that it can be considered a weak shock (Plooster 1968; Tsikulin 1970). Subsequent refinement of the cylindrical shock wave treatment was discussed by Hutchens (1995) where he gave the solutions to the cylindrical blast theory for the near-field scenario.

Relative to the shock wave generated by a typical hypersonic body (e.g. shock in front of a hypersonic body), the cylindrical shock wave strength is significantly reduced (Jones et al. 1968). The primary reason is that cylindrical shock waves are a function of energy deposited per unit length, and consequently temperature and pressure ratio ($p/p_0$), which is defined as the ratio between the meteor vapour pressure ($p$) and the ambient atmospheric pressure ($p_0$). For example, dissociation and excitation dominate in the region of strong meteor cylindrical shock wave propagation (Tsikulin 1970), while ionization occurs in small amounts of up to 1% in cylindrical shock waves with $M_{sw} \geq 28$ (Wilson 1966).

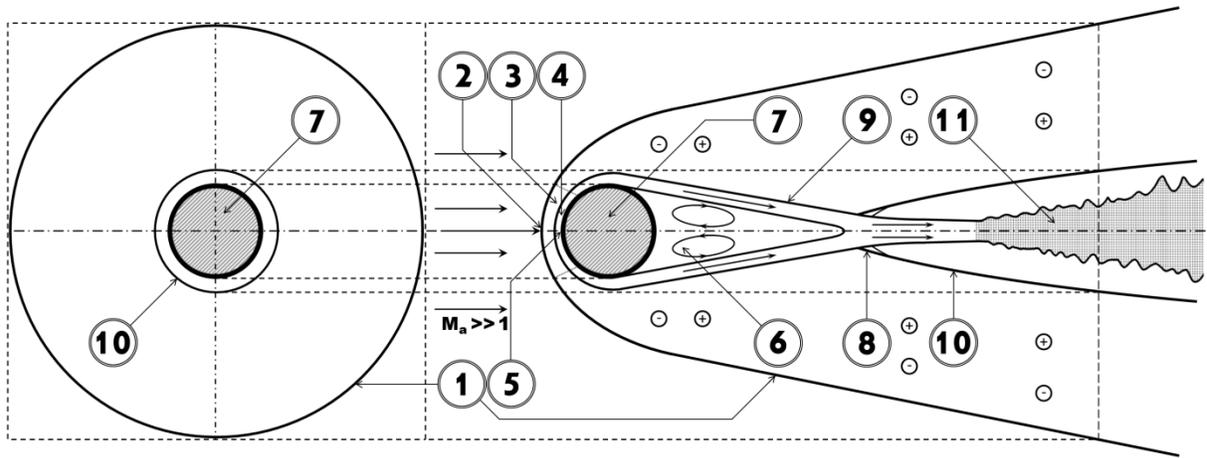

**Figure S1**: Schematics of the meteor shock wave(s), flow fields and near wake. The meteoroid is considered as a blunt body (with the spherical shape) propagating at hypersonic velocity. The definitions and explanations are provided in the text (after Hayes & Probstein (1959); Lees & Hromas (1961)). (1) Bow (cylindrical) shock wave front; (2) The "ballistic" shock front; (3) Sonic region; (4) Boundary layer; (5) Stagnation point; (6) Turbulent region (in some older literature, this is referred as the dead water region); (7) Meteoroid; (8) The neck and recompression region; (9) The 'free' shear layer; (10) The recompression vapour (or a true cylindrical) shock wave front; (11) The region of turbulent vapour flow and adiabatic expansion. Note that small circles with positive and negative signs indicate regions affected by the presence of ions and electrons respectively. The diagram is only for the illustrative purpose and is not to scale (e.g. for the reasons, see Jenniskens et al. 2000).